# Coarse-to-fine spatial GLMM

# for scalable prediction and multiscale analysis

Daisuke Murakami[1,2], Alexis Comber[3], Takahiro Yoshida[4],

Narumasa Tsutsumida[5], Chris Brunsdon[6], Tomoki Nakaya[7]

[1] Department of Fundamental Statistical Mathematics, Institute of Statitsical Mathematics, 10-3 Midoricho, Tachikawa, Tokyo, 190-8562 Japan

[2] Center for Urban & Real Estate Studies, Hitotsubashi University, Japan

[2] School of Geography, University of Leeds, UK

[4] Center for Spatial Information Science, the University of Tokyo, Japan

[5] Fujitsu Research, Fujitsu Limited, Kawasaki, Japan

[6] National Centre for Geocomputation, Maynooth University, Ireland

[7] Graduate School of Environment Studies, Tohoku University, Japan

Abstract: We develop a coarse-to-fine generalized linear mixed model (CF-GLMM) for scalable spatial prediction of exponential-family responses. CF-GLMM extends coarse-to-fine spatial modeling (CFSM) beyond Gaussian data by sequentially increasing spatial resolution while validation deviance improves, so the terminal resolution need not be specified in advance. The latent spatial process is constructed by aggregating local models, avoiding costly operations on large spatial covariance matrices. An idealized convolution representation links the covariance of our assumed spatial process to a sum of Matérn covariances with different ranges. Monte Carlo experiments show that CF-GLMM achieves predictive accuracy comparable to widely used scalable spatial GLMMs while requiring less computation and memory. Its scale-specific components also support exploratory multiscale analysis. An application to COVID-19 infection data in Tokyo illustrates fine-resolution spatial prediction and multiscale characterization of geographical variation.

# 1. Introduction

Spatial prediction is a central task in geographical analysis, with applications in environmental science, ecology, epidemiology, and urban studies. Many geographical outcomes are observed as counts, binary indicators, or other non-Gaussian variables. Spatial generalized linear mixed models (GLMMs; Diggle et al. 1998), which combine generalized linear models with latent spatial processes, provide a widely used framework for predicting such outcomes while accounting for spatial dependence. As increasingly large and fine-resolution geospatial datasets become available, however, spatial prediction must balance predictive accuracy with the computational cost required to represent sufficiently detailed spatial variation.

Gaussian processes (GPs; Cressie 1993; Williams and Rasmussen 2006) are commonly used to model latent spatial processes, but standard GP inference requires manipulating an $N \times N$ spatial covariance matrix, resulting in cubic computational complexity with the sample size $N$. Numerous scalable approximations have therefore been developed (see Heaton et al. 2019; Liu et al. 2020; Hazra et al. 2025 for reviews). Among them, basis-function approaches represent spatial variation using a finite set of functions, including kernels (Cressie and Johannesson 2008), splines (Wood 2017), and finite-element basis functions in the stochastic partial differential equation (SPDE) approach (Lindgren et al. 2011; Bakka et al. 2018). Such approximations substantially reduce computational cost, but their predictive performance depends on the complexity of the resulting approximate spatial representation. A coarse representation can oversmooth local structure relevant to prediction, whereas a finer representation, obtained by increasing the number of basis functions or mesh vertices, can capture finer-scale predictive structure at additional computational cost. Scalable spatial prediction therefore involves a trade-off among predictive accuracy, spatial resolution, and computational efficiency.

A practical difficulty is that the spatial resolution required for accurate prediction is generally unknown before model fitting. Mesh-refinement procedures (e.g., Righetto et al. 2020), smoothing-parameter selection (e.g., Wood 2017), multiresolution approximations (e.g., Nychka et al. 2015; Katzfuss2017), and full-scale or fused approximations (e.g., Sang and Huang 2012; Ma and Kang 2020) provide different ways of managing the trade-off between spatial resolution and computational cost. This issue is particularly important for large datasets, where increasingly fine-scale spatial variation may become predictable as sample size increases. From a predictive perspective, a natural strategy is therefore to start with a coarse representation and increase its resolution only as long as finer-scale structure improves out-of-sample predictive performance.

Coarse-to-fine spatial modeling (CFSM; Murakami et al. 2026) implements this idea for Gaussian spatial data. CFSM sequentially learns spatial components from coarse to fine scales and uses holdout validation to determine when additional refinement no longer improves prediction. Each scale-specific component is constructed by aggregating locally estimated models, thereby avoiding costly matrix operations involving a large spatial covariance matrix. The resulting model adapts its terminal spatial resolution to the predictive information supported by the data while retaining computational scalability. The original CFSM, however, is restricted to Gaussian responses and therefore cannot directly address the many geographical prediction problems involving counts, binary outcomes, and other exponential-family variables.

We develop a coarse-to-fine GLMM (CF-GLMM) that extends CFSM to exponential-family responses. CF-GLMM reformulates coarse-to-fine learning on the deviance scale using iteratively updated working responses and non-constant working weights. Spatial resolution is sequentially increased while validation deviance improves, so the terminal resolution need not be specified in advance. Because its latent spatial process remains an aggregation of local models, CF-GLMM also avoids costly operations on large covariance matrices. A further benefit of the coarse-to-fine construction is that the fitted spatial process is explicitly represented as a sum of scale-specific components. These components provide an exploratory decomposition of the spatial variation contributing to prediction.

The remainder of this paper is organized as follows. Section 2 introduces CFSM, and Section 3 develops CF-GLMM. Section 4 evaluates its predictive accuracy, uncertainty quantification, and computational efficiency through Monte Carlo experiments, while Section 5 examines its ability to recover spatial variation across scales. Section 6 applies CF-GLMM to COVID-19 infection data in Tokyo, and Section 7 concludes with a discussion.

## 2. Coarse-to-fine spatial modeling (CFSM)

This section introduces the Gaussian CFSM (Murakami et al. 2026). The CFSM considers a multiscale process that consists of scale-wise components $z_1(s_i), \ldots, z_R(s_i)$ corresponding to the bandwidth values $h_1, \ldots, h_R,$ where $h_r = \delta h_{r-1,}$ with $0 < \delta < 1$. In other words, $z_r(s_i)$ represents the $r$-th largest-scale process where $r \in \{1, \ldots, R\}$. The number of scales, *R, is assumed to be unknown*. Section 2.1 introduces the process model for $z_r(s_i)$, and Section 2.2 introduces the Gaussian regression model for the response variable.

## 2.1. Spatial process

The single-scale process $z_r(s_i)$ is constructed by aggregating $C_r$ local models distributed across the study area. Murakami et al. (2026) assumed that $C_r = round(1.5D^2/h_r^2)$ where $round(\cdot)$ denotes the rounding operator and *D* is the diagonal length of the bounding box containing the sample sites. The local centers, $s_{1_r}, \dots, s_{C_r}$, at which the local models are estimated, are determined as the *k-means cluster* centers.

The local model at the $c_r$-th center is defined given $h_r$, as follows:

$$z_r(s_i)|c_r = \mu_{c_r} + e_{c_r}, \mu_{c_r} \sim N(0, \tau_{c_r}^2), e_{c_r} \sim N\left(0, \frac{\sigma_{c_r}^2}{w_r^2(d_{i,c_r})}\right), \quad (1)$$

where $\mu_{c_r}$ and $\sigma_{c_r}^2$ denote the local mean and variance near the center. To estimate these parameters, the samples were locally weighted using $w_r(d_{i,c_r})$, a function that decays with the Euclidean distance $d_{i,c_r}$ from the center. We use an exponential kernel $w_r(d_{i,c_r}) = \exp\left(-\frac{d_{i,c_r}}{h_r}\right)$. The $C_r$ local models describe the local means and variances at each local center, given the bandwidth value $h_r$. $\tau_{c_r}^2$ is another variance parameter.

A single-scale process is constructed by aggregating these local models and taking the weighted product of their probability density functions following Cao and Fleet (2014). The resulting process is expressed as follows:

$$z_r(s_i) \sim N(\hat{z}_r(s_i), \hat{\sigma}_r^2(s_i)), \quad (2)$$

with mean $\hat{z}_r(s_i) = \sigma_r^2(s_i) \sum_{c_r=1}^{C_r} \frac{w_r(d_{i,c_r})}{\sigma_{c_r}^2} \mu_{c_r}$ and variance $\hat{\sigma}_r^2(s_i) = 1/\sum_{c_r=1}^{C_r} \frac{w_r(d_{i,c_r})}{\sigma_{c_r}^2}$.

CFSM cumulates the single-scale processes $z_1(s_i), \dots, z_R(s_i)$, whose scales are determined by the bandwidth values $h_1 > h_2, \dots, > h_R$, as follows:

$$z_{1:R}(s_i) = \sum_{r=1}^{R} z_r(s_i), \tag{3}$$

where $R$ is optimized by a validation method as detailed later.

## 2.2.Gaussian spatial regression

Murakami et al. (2026) assumed the following Gaussian model for response variable $y(s_i)$:

$$\begin{aligned} y(s_i) &= \mathbf{x}(s_i)'\boldsymbol{\beta} + z_{1:R}(s_i) + e(s_i), e(s_i) \sim N(0, v^2 w_y(s_i)), \\ z_{1:R}(s_i) &= \sum_{r=1}^{R} z_r(s_i), \quad z_r(s_i) \sim N(\hat{z}_r(s_i), \hat{\sigma}_r^2(s_i)), \end{aligned} \tag{4}$$

where $\mathbf{x}(s_i)$ is the covariate vector, $\boldsymbol{\beta}$ is the coefficient vector, and $v^2$ is the noise variance. $w_y(s_i)$ is the sample weight, which equals 1 in the original CFSM.

The model is optimized by holdout validation (HV), in which the samples are randomly divided into two subsets, with $100\alpha$ % assigned to a training set and $100(1-\alpha)$ % assigned to a validation set indexed by $i_v \in 1, \dots, N_v$, to minimize the following squared loss:

$$\sum_{i_v=1}^{N_v} w_y(s_{i_v}) \left( y(s_{i_v}) - \mathbf{x}(s_{i_v})'\boldsymbol{\beta} - \hat{z}_{1:R}(s_{i_v}) \right)^2. \tag{5}$$

Unlike conventional spatial statistical models, an explicit prior penalty in the objective function is unnecessary because HV implicitly regularizes and mitigates overfitting.

During the HV, the number $R$ of scales sequentially increases until the squared loss no longer decreases. Initially, $R = 1$ is assumed to estimate $\hat{z}_{1:R}(s_i) = \hat{z}_1(s_i)$ minimizing Eq. (5). If this reduces the squared loss, $R = 2$ is assumed to estimate $\hat{z}_{1:R}(s_i) = \hat{z}_1(s_i) + \hat{z}_2(s_i)$. Similarly, $R$ is gradually increased to train $\hat{z}_{1:R}(s_i) = \hat{z}_{1:R-1}(s_i) + z_R(s_i)$. In each step, $\hat{z}_R(s_i)$ is estimated as follows: (i) distribute the $C_R$ local centers using the $k$-means method; (ii) estimate the local model for each center (see Eq. (1)), and (iii) aggregate these local model estimates using Eq. (2).

# 3. CFSM-based spatial GLMM (CF-GLMM)

This section develops a CFSM-based spatial GLMM, which we will refer to as CF-GLMM. Section 3.1 introduces the model, followed by Section 3.2, which defines the deviance loss function minimized to optimize the model. Section 3.3 describes the optimization algorithm. Section 3.4 describes uncertainty quantification. Finally, Section 3.5 discusses properties of CF-GLMM.

## 3.1.Model

CF-GLMM assumes the following model:

$$y(s_i) \sim P(\mu(s_i), \varphi), \quad g(\mu(s_i)) = \mu_{lin}(s_i), \quad \mu_{lin}(s_i) = \mathbf{x}(s_i)'\boldsymbol{\beta} + o(s_i) + z_{1:R}(s_i),$$
$$z_{1:R}(s_i) = \sum_{r=1}^{R} z_r(s_i), \quad z_r(s_i) \sim N(\hat{z}_r(s_i), \hat{\sigma}_r^2(s_i)), \tag{6}$$

where $E[y(s_i)] = \mu(s_i)$ and $V[y(s_i)] = \varphi V[\mu(s_i)]$ where is the $\varphi$ dispersion parameter. $P(\mu(s_i))$ is an exponential family distribution with mean $\mu(s_i)$. $g(\cdot)$ is the link function, and $o(s_i)$ is a known offset variable. This model is identical to the conventional spatial GLMM, except that the multiscale process $z_{1:R}(s_i)$ is used instead of the spatial GP or its basis-function expression.

This study considers the following popular specifications: (a) a count data model with a Poisson distribution $P(\cdot) = Poisson(\cdot)$ and a log-link function $g(\mu) = \log(\mu)$; (b) a binary data model with a Bernoulli distribution $P(\cdot) = Bernoulli(\cdot)$ and a logistic-link function $g(\mu) = \log\frac{\mu}{1-\mu}$. Note that, if $P(\cdot)$ is a Gaussian distribution and $g(\mu(s_i)) = \mu(s_i)$, Eq. (6) can be reduced to the Gaussian CFSM (see Eq. (4)).

## 3.2.Deviance loss

Similar to the original CFSM, the proposed CF-GLMM (see Eq. (6)) is optimized via HV assuming $100\alpha$ % of training samples and $100(1-\alpha)$ % validation samples, where we assume $\alpha = 0.75$. HV minimizes the deviance loss for validation samples $y(s_{1_v}), \dots, y(s_{N_v})$. For Gaussian data, the squared loss in Eq. (5) corresponds to the deviance loss. For the Poisson model, the following Poisson loss is minimized:

$$Loss^{Pois} = 2\sum_{i_v=1}^{N_v}\left[y(s_{i_v})\log\left(\frac{y(s_{i_v})}{\mu(s_{i_v})}\right) - \left(y(s_{i_v}) - \mu(s_{i_v})\right)\right]. \quad (7)$$

For the binary model, the following logistic loss is minimized:

$$Loss^{Logit} = -2\sum_{i_v=1}^{N_v}\left[y(s_{i_v})\log\mu(s_{i_v}) + \left(1-y(s_{i_v})\right)\log\left(1-\mu(s_{i_v})\right)\right]. \quad (8)$$

Conventional spatial GLMMs typically regularize the latent spatial process through a covariance model or prior distribution. In contrast, CF-GLMM is optimized using the deviance loss, with model complexity controlled through HV. This is possible because our HV acts as an implicit regularization mechanism that mitigates overfitting, thereby eliminating the need for an explicit spatial regularization term in the objective function.

The second-order Taylor expansion of the validation deviance results in the following weighted squared loss:

$$\sum_{i_v=1}^{N_v}\hat{w}_y(s_{i_v})\left[\hat{\eta}(s_{i_v}) - \mathbf{x}(s_{i_v})'\boldsymbol{\beta} - o(s_{i_v}) - z_{1:R}(s_{i_v})\right]^2, \quad (9)$$

where $\hat{\eta}(s_{i_v}) = \hat{\mu}_{lin}(s_{i_v}) + \left(y(s_{i_v}) - \hat{\mu}(s_{i_v})\right)g'\left(\hat{\mu}(s_{i_v})\right)$ and $\hat{w}_y(s_{i_v}) = \frac{1}{V\left(\hat{\mu}(s_{i_v})\right)g'\left(\hat{\mu}(s_{i_v})\right)^2}$.

To minimize the deviance loss, the basic GLM alternately iterates the minimization of the weighted squared loss and updates the weight $\hat{w}_y(s_i)$ until convergence. However, this algorithm is not readily available because of the existence of $z_{1:R}(s_i)$. This difficulty is addressed by updating $\hat{w}_y(s_i)$ and $z_{1:R}(s_i)$ in each iteration and minimizing the weighted squared loss (see Eq. (9)). The details of the algorithm are described in the following subsection.

## 3.3.Optimization algorithm

To estimate the parameters $\boldsymbol{\beta}$, $R$, process means $\hat{z}_1(s_i), \ldots, \hat{z}_R(s_i)$ and variances $\hat{\sigma}_1^2(s_i), \ldots, \hat{\sigma}_R^2(s_i)$, we sequentially minimize Eq. (9), while increasing the number of scales $R$.

The HV begins at $R = 1$. $\hat{z}_{1:R}(s_i) = \hat{z}_1(s_i)$ and $\boldsymbol{\beta}$ in Eq. (6) are estimated using the training samples. The estimates are accepted if they reduce the validation loss, and $\hat{z}_{1:R}(s_i) = \hat{z}_1(s_i) = 0$ otherwise. Then, $R = 2$ is assumed, $\hat{z}_2(s_i)$ and $\boldsymbol{\beta}$ are estimated given $\hat{z}_{1:R-1}(s_i) = \hat{z}_1(s_i)$ using the training samples. These estimates are accepted if they reduce the validation loss, and $\hat{z}_2(s_i) = 0$ otherwise. The multiscale process is then updated as $\hat{z}_{1:R}(s_i) = \hat{z}_1(s_i) + \hat{z}_2(s_i)$. Thus, $\hat{z}_{1:R}(s_i)$ is sequentially updated until the validation loss no longer improves. In our implementation, the algorithm terminated if the loss did not improve for 5 consecutive $R$ values. The optimal $R$ value was chosen as the last value that improved the validation loss.

Specifically, after initializing $R = 1$, $Loss_{R-1} = \infty$, $\hat{z}_{1:R-1}(s_i) = 0$, $\hat{o}_{1:R-1}(s_i) = o(s_i)$ as well as $\hat{\boldsymbol{\beta}}$ by the basic GLM estimator, the HV-based learning algorithm is summarized as follows:

1. Form $\hat{\mu}_{lin}(s_i) = \mathbf{x}(s_i)'\hat{\boldsymbol{\beta}} + \hat{o}_{1:R-1}(s_i)$, $\hat{\mu}(s_i) = g^{-1}(\hat{\mu}_{lin}(s_i))$, and $\hat{w}_y(s_i) = \frac{1}{V(\hat{\mu}(s_i))g'(\hat{\mu}(s_i))^2}$

2. Given $\hat{z}_{1:R-1}(s_i)$, estimate $\hat{z}_R(s_i)$ and $\boldsymbol{\beta}$ to minimize the weighted squared loss for the training samples (see Eq. (9)):

$$\sum_{i_t=1}^{N_t} \hat{w}_y(s_{i_t}) \left[\hat{\eta}(s_{i_t}) - \mathbf{x}(s_{i_t})'\boldsymbol{\beta} - \hat{o}_{1:R-1}(s_{i_t}) - z_R(s_{i_t})\right]^2, \tag{10}$$

where $i_t \in \{1_t, \ldots, N_t\}$ represents an index for the training samples. $\hat{o}_{1:R-1}(s_{i_t}) = o(s_{i_t}) + \hat{z}_{1:R-1}(s_{i_t})$ is a given offset variable. Since Eq. (10) is identical to the loss function of the original

CFSM in the $R$-th step, it is estimated following Murakami et al. (2026) as follows:

2.1. Distribute $C_R$ local centers. Following the original CFSM, the centers are defined as the *k-means cluster* centers.

2.2. For each center $c_R \in \{1_R, \ldots, C_R\}$, fit the local model (see Eq. (1)) to estimate the local mean $\mu_{c_R}$ and variance $\sigma^2_{c_R}$.

2.3. Synthesize the $C_R$ estimated local models using Eq. (2), and specify the single-scale process $z_R(s_i) \sim N\left(\hat{z}_R(s_i), \hat{\sigma}^2_R(s_i)\right)$, where $\hat{z}_R(s_i) = \hat{\sigma}^2_R(s_i) \sum_{c_R=1}^{C_R} \frac{w_R(d_{i,c_R})}{\sigma^2_{c_R}} \mu_{c_R}$, and $\hat{\sigma}^2_R(s_i) = 1/\sum_{c_R=1}^{C_R} \frac{w_R(d_{i,c_R})}{\sigma^2_{c_R}}$.

2.4. Minimize Eq. (10) given $\hat{z}_{1:R-1}(s_i)$ and $\hat{z}_R(s_i)$ to obtain the weighted least squares estimate $\hat{\boldsymbol{\beta}}_R$ of $\boldsymbol{\beta}$.

3. Evaluate the validation loss $Loss_R$ of the model given $\hat{\boldsymbol{\beta}}_R$, $\hat{z}_{1:R-1}(s_{i_v})$, and $\hat{z}_R(s_{i_v})$:

(a) If $Loss_R < Loss_{R-1}$, $\hat{\boldsymbol{\beta}} = \hat{\boldsymbol{\beta}}_R$ and $\hat{o}_{1:R}(s_i) = \hat{o}_{1:R-1}(s_i) + \hat{z}_R(s_i)$, reset the counter $Q$=0, and go to Step 4.

(b) Otherwise, $Q \to Q + 1$. $\hat{o}_{1:R}(s_i) = \hat{o}_{1:R-1}(s_i)$. If $Q$ is less than a threshold value, which is 5 in our case, proceed to Step 4. Otherwise, $R$ is the terminal resolution, and $\hat{\boldsymbol{\beta}}$ and $\hat{z}_{1:R}(s_i) = \sum_{r=1}^{R} \hat{z}_r(s_i)$ are the outputs.

4. Update scale $R \to R + 1$, reduce the bandwidth $h_{R+1} = \delta h_R$ (see Section 2), where we assumed $\delta = 0.9$, and go back to Step 1.

In short, this algorithm sequentially estimates $\hat{z}_1(s_{i_v}),\dots,\hat{z}_R(s_{i_v})$ until the deviance loss no longer improves. Owing to Step 3, this algorithm never increases the deviance loss over the iterations.

## 3.4.Uncertainty quantification

This section summarizes our uncertainty quantification. See Appendix 1 for further details. For computational simplicity, we assumed independence among $z_1(s_i),\dots,z_R(s_i)$, so that the variance of the multiscale process is

$$\hat{\sigma}_{1:R}^2(s_i) = \sum_{r=1}^{R} \hat{\sigma}_r^2(s_i). \tag{11}$$

Although this assumption may be reasonable because each component $z_r(s_i)$ is trained on residual variation unexplained by coarser-scale components $z_1(s_i),\dots,z_{r-1}(s_i)$, it may still bias the variance estimate. We therefore calibrate $\hat{\sigma}_{1:R}^2(s_i)$ using the validation samples. Specifically, we rescale the model-based variance $\hat{\sigma}_{1:R}^2(s_i)$ so that its average matches the empirical process variance $\hat{\sigma}_{emp}^2$ estimated from the validation samples. The rescaled process variance is denoted as $\hat{\sigma}_{1:R}^{*2}(s_i)$.

Using $\hat{\sigma}_{1:R}^{*2}(s_i)$, the variance of the predictive mean is obtained by the delta method, as follows:

$$Var[\hat{\mu}(s_i)] = \left(\frac{d\mu(s_i)}{d\mu_{lin}(s_i)}\right)^2 \left(\mathbf{x}(s_i)'\hat{\mathbf{\Sigma}}_{\boldsymbol{\beta}}\mathbf{x}(s_i) + \hat{\sigma}_{1:R}^{*2}(s_i)\right), \tag{12}$$

where $\mathbf{x}(s_i)$ is the vector of covariates observed at site $s_i$. $\hat{\mathbf{\Sigma}}_{\boldsymbol{\beta}}$ is the covariance matrix of the regression coefficients $\hat{\boldsymbol{\beta}}$. Because $\hat{z}_{1:R}(s_i)$ is treated as known when estimating $\hat{\boldsymbol{\beta}}$, the naïve model-based approach underestimates its variance. We therefore use a spatial-block cluster-robust estimator (Bester et al. 2011) to estimate $\widehat{\boldsymbol{\Sigma}}_{\boldsymbol{\beta}}$ as detailed in Appendix 1. The variance of the *k*-th coefficient estimate is evaluated as $Var[\hat{\beta}_k] = [\widehat{\boldsymbol{\Sigma}}_{\boldsymbol{\beta}}]_{kk}$, which is the *k*-th diagonal element of the $\widehat{\boldsymbol{\Sigma}}_{\boldsymbol{\beta}}$ matrix.

## 3.5.Properties

By expanding $z_{1:R}(s_i)$ based on Eqs. (1) and (2), the spatial process can be expressed as a basis-function model:

$$z_{1:R}(s_i) = \sum_{r=1}^{R}\sum_{c_r=1}^{C_r} \widetilde{w}_r(d_{i,c_r})\gamma_{c_r} + u(s_i), \qquad u(s_i) \sim N(0, \hat{\sigma}_{1:R}^2(s_i)). \tag{13}$$

The basis function $\widetilde{w}_r(d_{i,c_r}) = \hat{\sigma}_r^2(s_i) w_r(d_{i,c_r})$ is given by the kernel $w_r(d_{i,c_r})$ adjusted by the variance of the corresponding local model $\sigma_{c_r}^2$. Its coefficient estimate $\gamma_{c_r} = \frac{\mu_{c_r}}{\sigma_{c_r}^2}$ is determined by the local model parameters. The total number of basis functions $L = \sum_{r=1}^{R} C_r$ is determined by the HV-based selection of $R$ (see Section 3.3). Thus, within the proposed construction, the terminal $L$ need not be specified a priori, although the bandwidth sequence and other design choices remain fixed. This validation-guided refinement adaptively increases spatial resolution when adding a finer-scale component improves validation performance. Furthermore, $L$ can exceed $N$ when further refinement continues to improve validation performance; this is because local models are estimated sequentially rather than jointly. Thus, CF-GLMM provides a flexible representation of spatial processes that can accommodate both simple and complex small-scale patterns.

CF-GLMM with our assumed exponential kernel admits a covariance interpretation under an idealized process-convolution model (Higdon 2002). Under Gaussian white-noise measures across scales and convolution over the spatial domain, each scale-wise process has a covariance given by the kernel autoconvolution, yielding a Matérn covariance with smoothness $\nu = 2$ and a range proportional to the bandwidth (Stein 1999; Williams and Rasmussen 2006). As a result, the idealized multiscale process of CF-GLMM has a covariance equal to a sum of Matérn covariances with different ranges. This idealized model serves only as a theoretical reference; it does not imply that CF-GLMM estimation is equivalent to Matérn GP inference. See Appendix 2 for further details.

Computational efficiency is another important property. Unlike conventional spatial GP/GLMMs based on covariance modeling, CF-GLMM represents the spatial process as an aggregation of local models. Standard dense GP covariance modeling requires the inversion of covariance matrices, resulting in computational complexity of $O(N^3)$. In contrast, the computational complexity of our local model aggregation equals $O(N)$ per scale, or $O(N \log N)$ over the logarithmically growing number of coarse-to-fine scales (see Appendix 3), and is naturally amenable to parallel implementation, thereby achieving substantially improved scalability for large samples. The per-scale linearity reflects a coarse-to-fine balance: coarse scales fit few local models over many neighbors, whereas fine scales fit many local models over few neighbors, so the number of local models times the neighbors per model stays proportional to $N$.

The model is optimized using HV, whereas conventional spatial GLMMs are typically estimated using Bayesian or likelihood-based inference. Although likelihood-based models can also be integrated into modern machine-learning pipelines, the validation-based optimization of CF-GLMM provides a model-agnostic interface that requires only observed responses and out-of-sample predictions. The same coarse-to-fine selection rule can therefore be used when the resulting scale-wise components are incorporated into predictive algorithms such as neural networks and random forests, without constructing a joint spatial likelihood, as demonstrated by Murakami et al. (2026).

The aforementioned properties indicate flexibility, computational efficiency, and high extensibility. The performance of the proposed method is verified in the following sections.

# 4. Monte Carlo experiments 1: Spatial prediction

Sections 4 and 5 present Monte Carlo experiments that investigate the performance of the proposed method in terms of predictive accuracy and multiscale feature extraction. The main experiments focus on count data, with Gaussian and binary cases also reported in Section 4.5. Supplementary Material 1 examines the binary case in more detail.

## 4.1.Outline

This section evaluates the predictive accuracy of our method using synthetic count data given at locations $s_i \in \{s_1, \dots, s_N\}$ randomly distributed within the region [0, 10] × [0, 10]. A sample was generated for each site as follows:

$$y(s_i) \sim Poisson(\mu(s_i)), \qquad \log \mu(s_i) = \beta_0 + \beta_1 x_1(s_i) + \beta_2 x_2(s_i) + z(s_i),$$
$$z(s_i) = \sum_{j=1}^{N} w_h^{scale}(d_{ij})\, u(s_j), \qquad u(s_j) \sim N(0, 2^2), \tag{14}$$

where $\{\beta_1, \beta_2\} = \{2.0, -0.5\}$. $w_h^{scale}(d_{ij}) = \frac{w_h(d_{ij})}{\sum_j w_h(d_{ij})}$ is a scaled weight where $w_h(d_{ij}) = \exp(-d_{ij}/h)$. The bandwidth $h$ is defined as the average distance to the 10 nearest neighbors. This specification assumes an approximately constant sampling density within the bandwidth distance irrespective of the sample size, implying a smaller bandwidth for greater *N*. As shown in Figure 1, the resulting spatial process $z(s_i)$ exhibits smaller-scale patterns for larger *N*, making it suitable for evaluating scalability in terms of predictive accuracy.

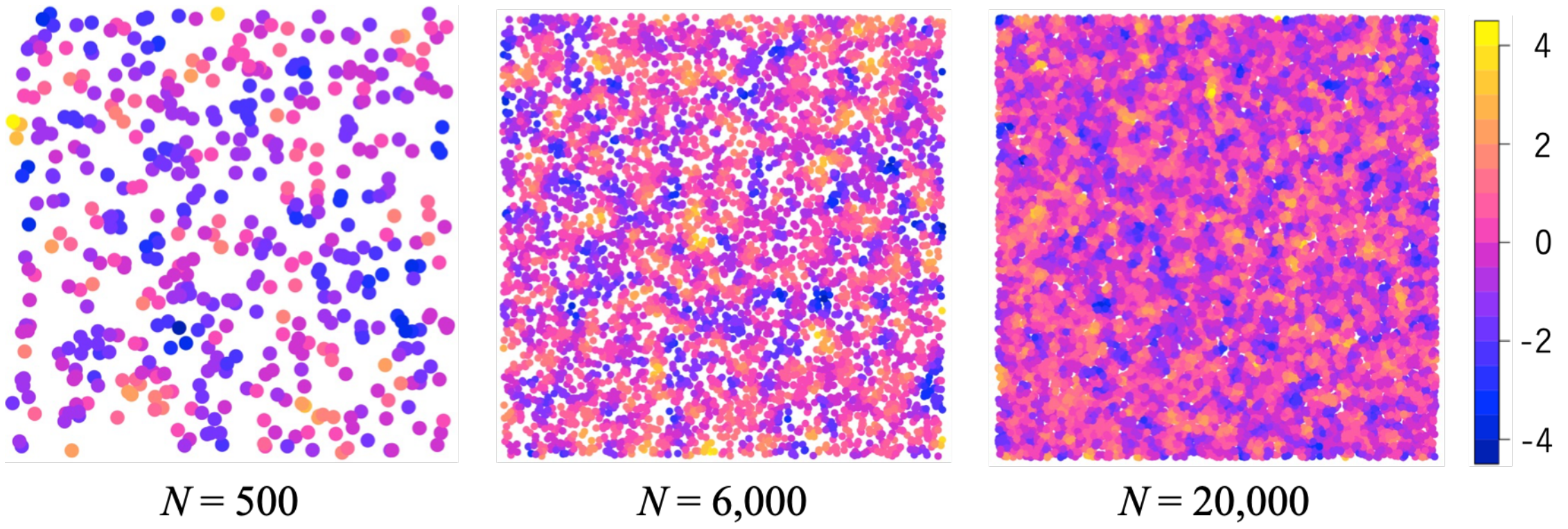

Figure 1: Spatial process $z(s_i)$ simulated using the second line of Eq. (14).

Spatial processes were also assumed in the covariables as follows:

$$x_k(s_i) = 0.5z_k(s_i) + 0.5e_k(s_i), e_k(s_i) \sim N(0,1),$$
$$z_k(s_i) = \sum_{j=1}^{N} w_h^{scale}(d_{ij})\, u_k(s_j), \qquad u_k(s_j) \sim N(0,1). \tag{15}$$

The simulated samples were generated 200 times for each case with intercept $\beta_0 \in \{-1.5, 0.5\}$ and training sample sizes $N \in \{500,1000,2000,3000,6000,12000,20000\}$. $\beta_0 = -1.5$ implies a many-zero scenario in which the probability of observing $y(s_i) = 0$ equals 0.800, and $\beta_0 = 0.5$ implies a few-zero scenario in which the probability of observing $y(s_i) = 0$ equals 0.192.

In each trial, 2,000 test samples were generated using Eqs. (14) and (15) and used to evaluate the mean absolute error (MAE), which quantifies point prediction accuracy, and the continuous ranked probability score (CRPS), which evaluates the overall quality of probabilistic predictions by accounting for both the location and spread of the predictive distribution.

The MAE and CRPS values of our CF-GLMM were compared with those of three Poisson benchmarks: a basic Poisson GLM (GLM), GAMs, and SPDE models. Among scalable spatial GLMMs, GAM and SPDE models are adopted because they are widely used and off-the-shelf software packages are available. The GAMs were implemented using the R package mgcv (version 1.9.4; https://cran.r-project.org/web/packages/mgcv/index.html), with basis functions specified to approximate a Gaussian process following Comber et al. (2024). We considered five GAM specifications with 33 (the default setting in mgcv), 100, 200, 400, and 800 basis functions, and the approximation proposed by Li and Wood (2020) was used to enhance scalability.

The SPDE models were implemented using the INLA package (version 25.10.19; https://www.r-inla.org/home). We considered six mesh specifications with the cutoff argument set to 1.0, 0.5, 0.2, 0.1, 0.05, and 0.02. The cutoff argument controls the minimum allowed distance between mesh vertices, with smaller values generally producing denser meshes and more basis functions.

All computations were performed using R 4.6.0 on a Mac Studio equipped with an Apple M3 Ultra chip and 512 GB of unified memory.

## 4.2.Result: Predictive accuracy

Figure 2 compares the predictive MAE values. The basic GLM performed poorly because it ignored latent spatial processes. GAMs with fewer basis functions yielded larger MAEs. Although the GAM with 800 basis functions achieved reasonably small MAEs for small samples, the MAE deteriorated as $N$ increased. The SPDE corresponding to cutoff > 0.1 suffered from a similar degeneracy problem. By contrast, SPDE was highly accurate when the cutoff value was equal to or smaller than 0.1, confirming the necessity of careful tuning of the mesh density.

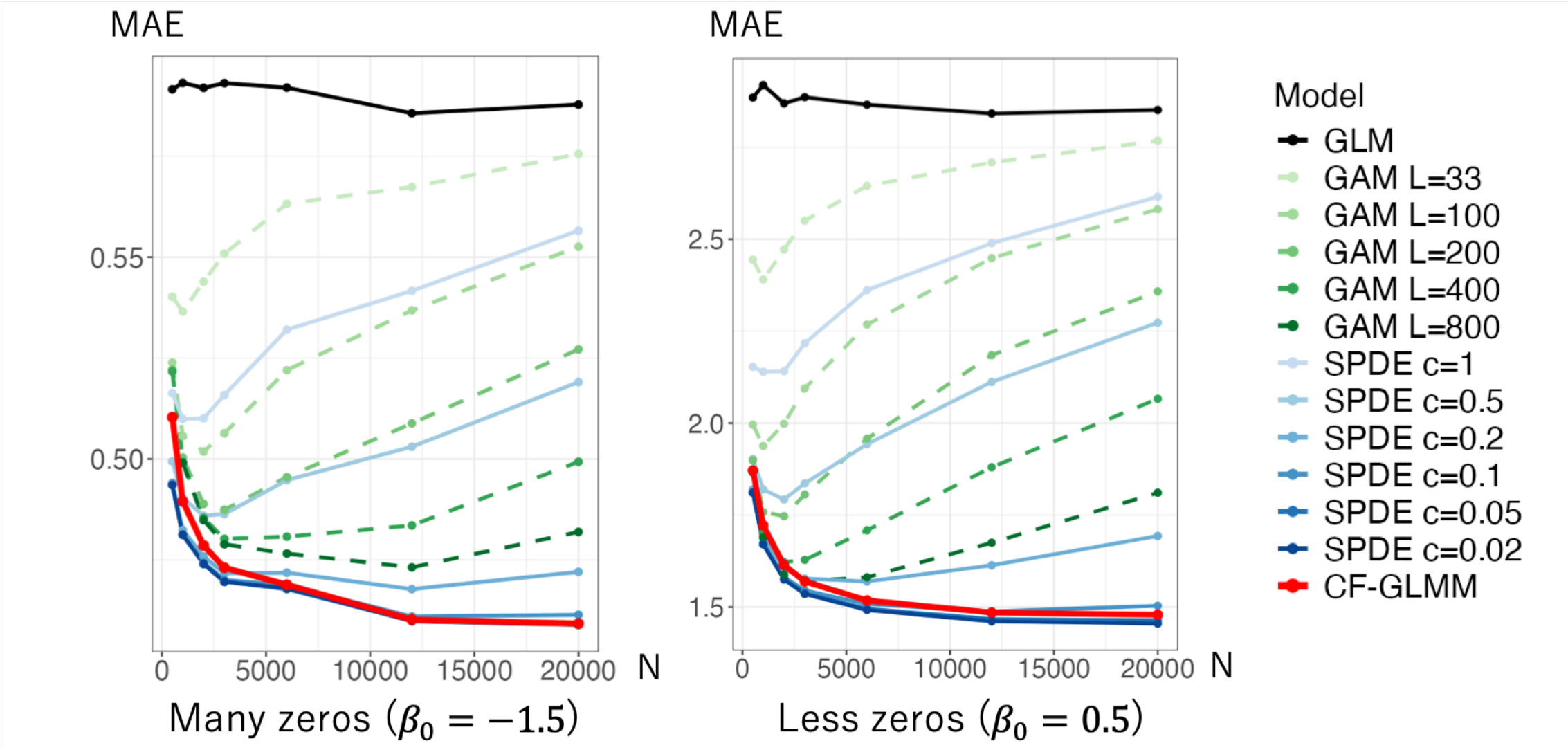

Figure 2: Predictive MAEs. "c" in the legend denotes the cutoff parameter.

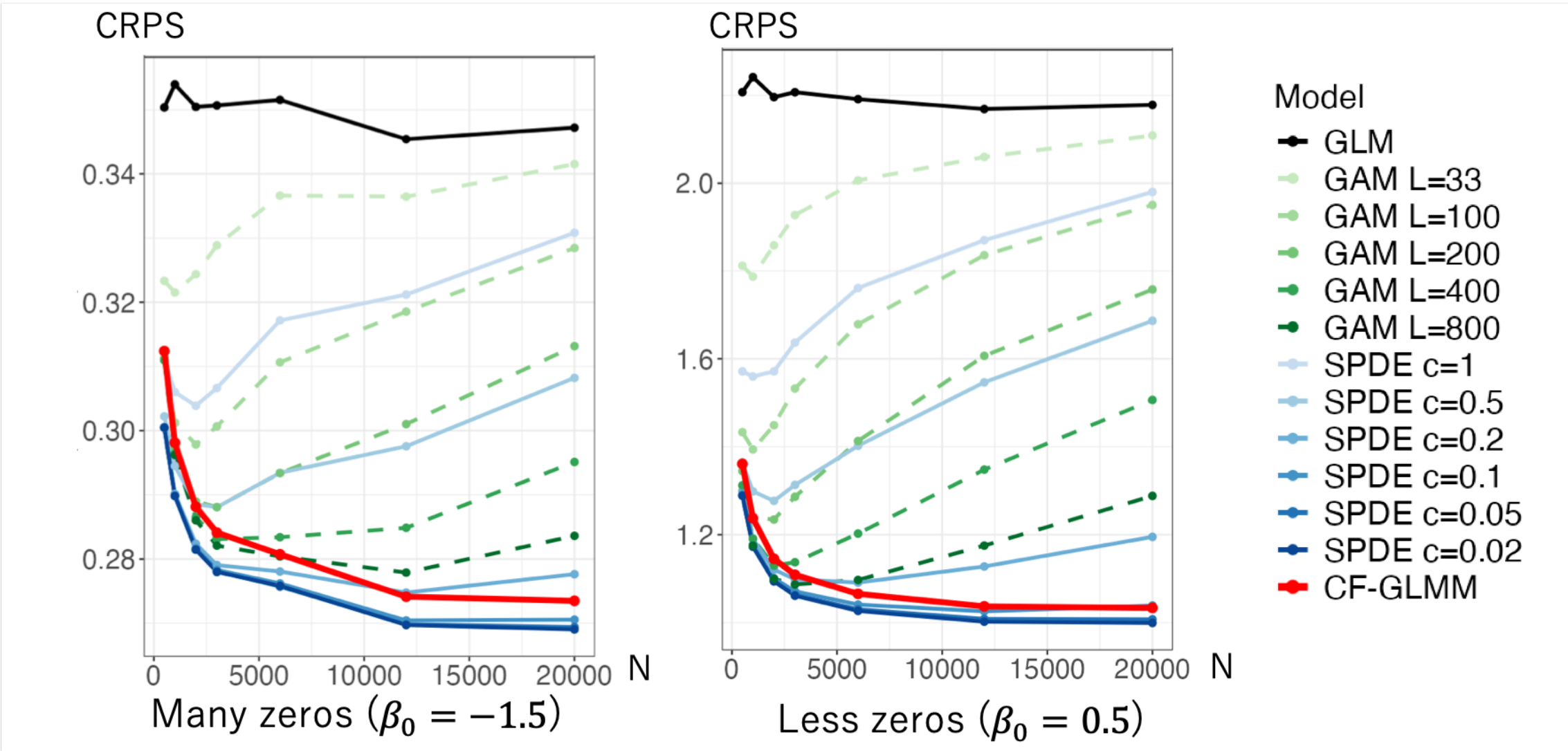

Figure 3: Predictive CRPS.

The MAE of CF-GLMM decreased as $N$ increased and became comparable to that of the SPDE with small cutoff values, suggesting that CF-GLMM captures both large- and small-scale spatial processes reasonably well. Importantly, this accuracy was attained without requiring the terminal number of basis functions $L$ to be specified in advance.

Figure 3 compares the CRPS values. CF-GLMM yielded slightly worse CRPS values than the SPDEs with small cutoff values, particularly under the many-zero scenario ($\beta_0 = -1.5$). This may be because the SPDE places an explicit global Gaussian prior on the spatial process, which stabilizes the predictive distribution, whereas CF-GLMM regularizes only locally, through the Gaussian prior in each local model and implicitly through the HV. Nevertheless, the differences in CRPS were small. A similar tendency was found for the binary case in Supplementary Material, where the AUC of CF-GLMM was slightly lower than that of the SPDEs with small cutoff values. These results suggest that there is room to stabilize CF-GLMM for responses that carry limited information per observation, such as counts with many zeros and binary outcomes.

Importantly, all the spatial models discussed above performed well when the latent process was coarse. To see this, the true bandwidth was replaced with $h$ = 2.0, and the same Monte Carlo experiment with $\beta_0 = 0.5$ was performed. Figure 4 summarizes the resulting MAE and CRPS values; the GAM, SPDE, and CF-GLMM specifications performed similarly.

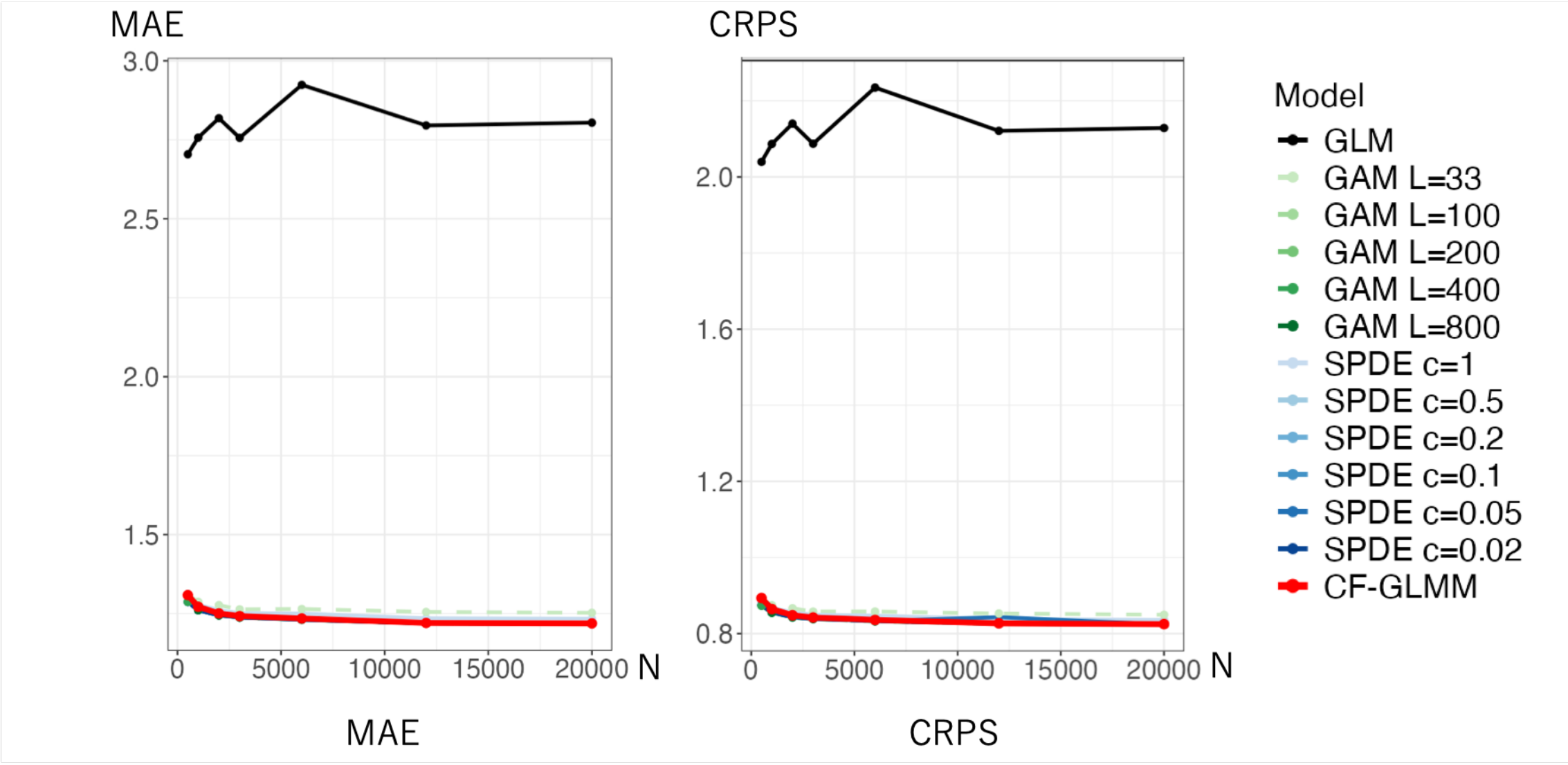

Figure 4: MAEs and CRPS in cases with coarser-scale spatial process ($h = 2.0$; $\beta_0 = 0.5$).

## 4.3.Result: Coefficient estimates

Figure 5 shows boxplots of the estimated $\beta_1$ values. The GLM estimate was upwardly biased because it ignored the latent process. The estimates of GAMs with few basis functions and SPDEs with large cutoff values were also upwardly biased, owing to insufficient flexibility to represent the assumed spatial process. The bias of CF-GLMM was the smallest owing to the accurate spatial process modeling, confirming its usefulness for regression analysis in the presence of spatial dependence (see LeSage and Pace 2009).

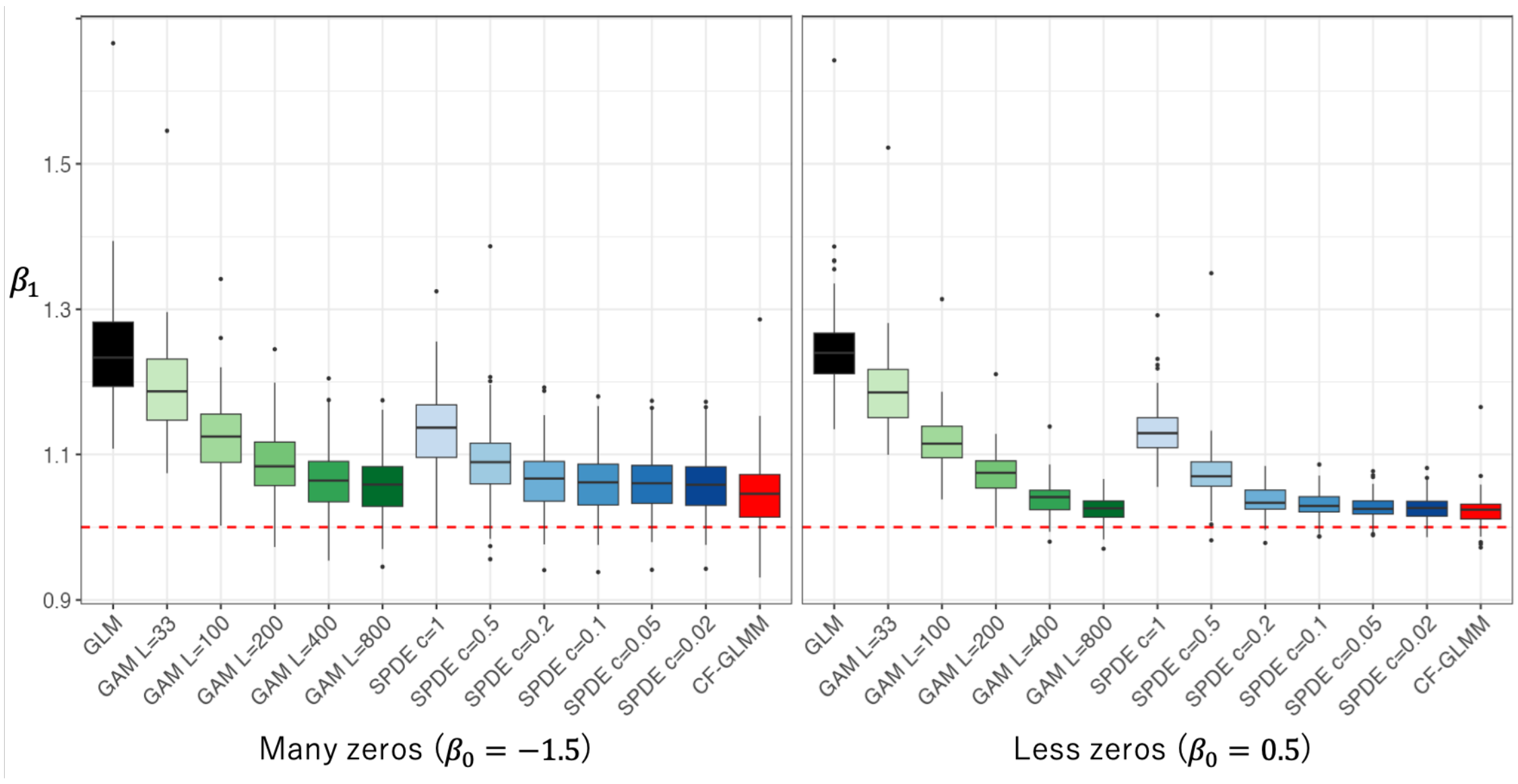

Figure 5: Boxplot of the coefficient estimate $\hat{\beta}_1$ ($N$ = 6,000). The red lines denote the true value.

Table 1: Mean SE and 95 % coverage probability of $\hat{\beta}_1$.

| Bandwidth $h$ | model | Gaussian | | Poisson | | Binary | |
|---|---|---|---|---|---|---|---|
| | | Mean SE | 95 % coverage | Mean SE | 95 % coverage | Mean SE | 95 % coverage |
| 10 nearest-neighbor distance | GLM | 0.036 | 0.00 | 0.014 | 0.00 | 0.057 | 0.82 |
| | GAM $L$=800 | 0.028 | 0.57 | 0.016 | 0.60 | 0.061 | 0.89 |
| | SPDE $c$=0.2 | 0.028 | 0.62 | 0.016 | 0.49 | 0.062 | 0.69 |
| | SPDE $c$=0.02 | 0.028 | 0.69 | 0.019 | 0.77 | 0.063 | 0.64 |
| | CF-GLMM | 0.030 | 0.77 | 0.028 | 0.93 | 0.064 | 0.93 |
| 2.0 | GLM | 0.037 | 0.92 | 0.014 | 0.39 | 0.056 | 0.27 |
| | GAM $L$=800 | 0.026 | 0.96 | 0.015 | 0.96 | 0.062 | 0.95 |
| | SPDE $c$=0.2 | 0.026 | 0.96 | 0.016 | 0.98 | 0.064 | 0.96 |
| | SPDE $c$=0.02 | 0.026 | 0.96 | 0.018 | 0.98 | 0.069 | 0.90 |
| | CF-GLMM | 0.029 | 0.98 | 0.017 | 0.96 | 0.065 | 0.95 |

Table 1 summarizes the coefficient standard error (SE) of $\hat{\beta}_1$ estimated following Section 3.4 and Appendix 1, together with the corresponding 95 % coverage probability. The SE of CF-GLMM tended to be greater than that of the other spatial models. While all the spatial models attained coverage close to the nominal 95 % level when the spatial process was coarse ($h$ = 2.0), CF-GLMM yielded coverage closest to the nominal level when the process contained fine-scale variation (h = 10 nearest-neighbor distance), suggesting better calibration of coefficient uncertainty in this setting.

## 4.4.Result: Computational cost

Table 2 compares average computation times of the GAM with 800 basis functions and SPDEs with cutoff = 0.2, 0.1, 0.05, and 0.02, which achieved MAE or CRPS values comparable to those of CF-GLMM. As a benchmark, the conventional full Gaussian process, whose computational complexity rapidly grows on the order of $N^3$, was also estimated using the spaMM package (https://cran.r-project.org/web/packages/spaMM/index.html).

The GAM, SPDEs, and CF-GLMM were considerably faster than the full Gaussian process, demonstrating favorable scalability over the sample sizes considered. The SPDE and CF-GLMM specifications tended to be faster than the GAM with $L$ = 800. When the comparison was restricted to specifications whose median MAE was within 97–103 % of that of CF-GLMM (shown in bold in Table 2), CF-GLMM was the fastest for most sample sizes. These results confirmed the computational efficiency of CF-GLMM.

Table 2: Comparison of computation time (seconds) across models whose MAE or CRPS values are comparable to those of CF-GLMM ($\beta_0 = 0.5$). Bold indicates cases where the MAE is within 97–103 % of that of CF-GLMM, whereas bold italics indicate cases where the MAE is more than 3 % lower than that of CF-GLMM.

| N | GP | GAM | SPDE | | | | CF-GLMM |
|---|---|---|---|---|---|---|---|
| | | $L$ = 800 | cutoff=0.20 | cutoff=0.10 | cutoff=0.05 | cutoff=0.02 | |
| 500 | **17.50** | [1] | **1.73** | **1.89** | **1.99** | **2.07** | 0.46 |
| 1,000 | ***140.57*** | **19.8** | **1.96** | ***2.37*** | ***2.66*** | ***2.87*** | 0.66 |
| 2,000 | **898.65** | **27.5** | **2.33** | **3.20** | **4.03** | **4.66** | 1.17 |
| 3,000 | **3,193.91** | **34.3** | **2.62** | **3.90** | **5.30** | **6.51** | 1.72 |
| 6,000 | [2] | 50.1 | 3.09 | **5.37** | **8.64** | **11.9** | 3.45 |
| 12,000 | | 43.7 | 4.05 | **7.45** | **14.8** | **24.5** | 7.54 |
| 20,000 | | 50.8 | 4.94 | **9.44** | **20.2** | **39.6** | 13.5 |

[1]Could not be computed because the number of basis functions exceeded $N$, [2] Too slow.

Table 3 compares the peak resident set size (RSS), quantifying memory usage. CF-GLMM had the smallest peak RSS across all cases, suggesting its potential for very large samples that would otherwise require substantial memory.

Table 3: Comparison of peak RSS (MB) across models whose MAE or CRPS values are comparable to those of CF-GLMM ($\beta_0 = 0.5$).

| N | GP | GAM | SPDE | | | | CF-GLMM |
|---|---|---|---|---|---|---|---|
| | | $L$=800 | cutoff=0.20 | cutoff=0.10 | cutoff=0.05 | cutoff=0.02 | |
| 500 | 564 | [1] | 435 | 472 | 479 | 482 | 304 |
| 1,000 | 936 | 586 | 464 | 528 | 524 | 541 | 310 |
| 2,000 | 2,444 | 930 | 522 | 613 | 665 | 713 | 327 |
| 3,000 | 5,737 | 973 | 584 | 664 | 753 | 865 | 333 |
| 6,000 | [2] | 972 | 670 | 762 | 956 | 1,101 | 375 |
| 12,000 | | 1,032 | 669 | 818 | 1,057 | 1,213 | 396 |
| 20,000 | | 1,328 | 825 | 975 | 1,306 | 1,672 | 509 |

[1] Could not be computed because the number of basis functions exceeded $N$, [2] Computationally infeasible in our environment.

## 4.5.Experiments on Gaussian, Poisson, and binary responses

We performed additional Monte Carlo experiments for Gaussian $y(s_i)\sim N(\mu(s_i), 1^2)$ and binary responses $y(s_i)\sim Bernoulli(\mu(s_i))$. The other settings followed those in Section 4.1. Figure 6 compares the resulting MAEs and computation times for the Gaussian case, as well as for the Poisson case. For the binary case, where MAE is less suitable, the area under the receiver operating characteristic curve (AUC) was evaluated using the predicted values and subtracted from one to obtain the AUC loss, which, like the MAE, takes smaller values for more accurate models. Figure 6 highlights the efficiency of CF-GLMM in achieving accurate spatial prediction with shorter computation time across the three data distributions.

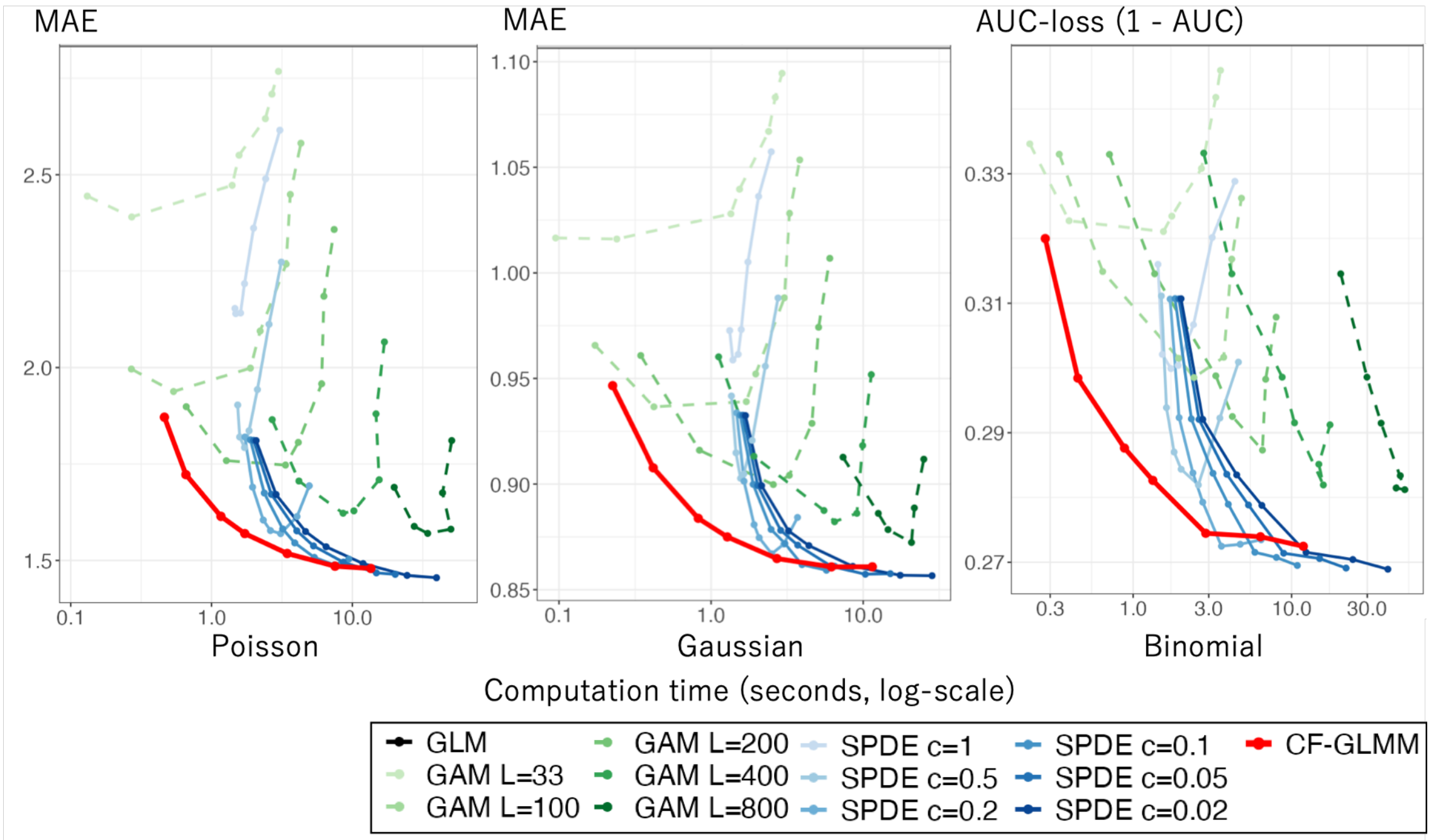

Figure 6: MAE (AUC loss for the binary case) and computation time across the three data distributions.

# 5. Monte Carlo experiments 2: Multiscale analysis

## 5.1. Outline

In this section, we evaluate the performance of the proposed method in terms of multiscale spatial feature extraction. The same count data generation process as in Section 4.1 was assumed, except that the spatial process $z(s_i)$ was replaced with the following multiscale process:

$$z(s_i) = \sum_{m=1}^{3} Z_m(s_i), \quad Z_m(s_i) = \sum_{j=1}^{N} w_{h_m}^{scale}(d_{ij})\, u_m(s_j), \quad u_m(s_j) \sim N(0,1). \qquad (16)$$

$w_{h_m}^{scale}(d_{ij}) = \frac{w_{h_m}(d_{ij})}{\sum_j w_{h_m}(d_{ij})}$ where $w_{h_m}(d_{ij}) = \exp(-d_{ij}/h_m)$ . The bandwidths for $Z_1(s_i), Z_2(s_i), Z_3(s_i)$ were $h_1 = 3.0$, $h_2 = 0.8$, $h_3 = 0.3$ assuming large-, moderate-, and small-scale spatial patterns, respectively.

In this section, we examine whether these three components can be recovered. For recovery, we fit the Poisson CF-GLMM. The predictive mean is then decomposed into three scales, $\hat{z}_{1:R}(s_i) = \hat{Z}_1(s_i) + \hat{Z}_2(s_i) + \hat{Z}_3(s_i)$ based on the intermediate value of the three bandwidths:

$$\hat{Z}_1(s_i) = \sum_{1.9 \le h_r} \hat{z}_r(s_i), \quad \hat{Z}_2(s_i) = \sum_{0.5 \le h_r < 1.9} \hat{z}_r(s_i), \quad \hat{Z}_3(s_i) = \sum_{h_r < 0.5} \hat{z}_r(s_i). \tag{17}$$

The accuracy of the estimated components was evaluated using correlation coefficients. Because scale-wise components are typically interpreted through mapping, the correlation coefficient was used to measure their similarity to the true patterns.

Instead of GLM and SPDE, which are not readily available for multiscale feature extraction, the multiscale recovery performance was compared with that of Moran eigenvector maps (MEMs), a multiscale spatial analysis method widely used in ecology (Dray et al. 2006; Legendre and Legendre 2012). MEMs represent the spatial process $\mathbf{z}^{MEM} = [z^{MEM}(s_1), \ldots, z^{MEM}(s_N)]'$ as $\mathbf{z}^{MEM} = \sum_{l=1}^{L} \mathbf{e}_l \gamma_l$ where $\mathbf{e}_1, \ldots, \mathbf{e}_L$ are the eigenvectors corresponding to positive eigenvalues $\lambda_1 \ge \lambda_2 \ge, \ldots, \lambda_L > 0$, of a matrix derived by doubly centering a spatial proximity matrix **W**. In our case, the (*i*, *j*)-th element is $w_{h_{mst}}(d_{ij}) = \exp(-d_{ij}/h_{mst})$ where $h_{mst}$ is the maximum edge length of the minimum spanning tree connecting the sample sites. As detailed by Dray et al. (2006), $\mathbf{e}_1, \ldots, \mathbf{e}_L$ act as multiscale basis functions, where $\mathbf{e}_l$ explains the *l*-th largest-scale spatial pattern. The eigenvectors are treated as additional covariates in the Poisson regression and selected using a forward stepwise method that minimizes the Poisson deviance.

We compare CF-GLMM with an ideal MEM that decomposes the estimated spatial process $\hat{\mathbf{z}}^{MEM} = \sum_{l=1}^{L} \mathbf{e}_l \hat{\gamma}_l$ into the three scales by maximizing the correlation coefficient with the true values $Z_m(s_i)$, as follows:

(1) Optimize the number $L_1$ of eigenvectors in $\hat{\mathbf{Z}}_1^{(MEM)} = \sum_{l=1}^{L_1} \mathbf{e}_l \hat{\gamma}_l$ by maximizing the correlation coefficient between $\hat{\mathbf{Z}}_1^{(MEM)}$ and the true values $\mathbf{Z}_1 = [Z_1(s_1), \dots, Z_1(s_N)]'$.

(2) Optimize $L_2$ in $\hat{\mathbf{Z}}_2^{(MEM)} = \sum_{l=L_1+1}^{L_2} \mathbf{e}_l \hat{\gamma}_l$ in the same manner.

(3) Evaluate $\hat{\mathbf{Z}}_3^{(MEM)} = \sum_{l=L_2+1}^{L} \mathbf{e}_l \hat{\gamma}_l$.

If the three components extracted from CF-GLMM indicate higher correlation coefficients than the ideal MEM, CF-GLMM is a promising alternative for multiscale spatial analysis.

Because the MEM assumes in-sample prediction, the correlation coefficients are evaluated using in-sample predictions of the factors with $N = 3{,}000$ samples. However, CF-GLMM is available for out-of-sample prediction of the components, as demonstrated in Section 6.

## 5.2.Result

Table 4 compares the correlation coefficients for each scale, and Figure 7 plots the predicted components spatially for one trial. The CF-GLMM tends to achieve higher correlations across scales, even though the MEM decomposition was optimized using the true scale-specific components. This is attributable to the larger number of basis functions (local models) in CF-GLMM; on average, MEMs and CF-GLMM considered 570 and 13,715 basis functions/eigenvectors, respectively. The CF-GLMM indicated considerably higher correlation values for the small-scale process, owing to the learning scheme that sequentially learns toward finer spatial scales until convergence, in contrast to MEMs, for which the number of eigenvectors is fixed a priori.

These results clearly suggest that CF-GLMM is useful for ecological and other analyses aimed at extracting multiscale features. In addition, our results suggest the potential of CF-GLMM as an efficient extractor of multiscale features used in machine-learning tasks (Kopczewska 2022). Still, the correlation values tend to be low at smaller scales, mainly due to the complexity of the spatial patterns and the difficulty of distinguishing small-scale processes from independent noise, which can be regarded as a smallest-scale process. Smaller-scale features should be interpreted with caution.

Table 4: Summary statistics of the correlation coefficients

| Scenario | Statistics | Large-scale | | Moderate-scale | | Small-scale | |
|---|---|---|---|---|---|---|---|
| | | CF-GLMM | MEM | CF-GLMM | MEM | CF-GLMM | MEM |
| Many-zero ($\beta_0 = -1.5$) | Max | 0.917 | 0.861 | 0.706 | 0.716 | 0.620 | 0.644 |
| | 75-percentile | 0.802 | 0.720 | 0.525 | 0.544 | 0.483 | 0.359 |
| | Median | 0.740 | 0.657 | 0.451 | 0.412 | 0.442 | 0.277 |
| | 25-percentile | 0.615 | 0.556 | 0.378 | 0.211 | 0.380 | 0.141 |
| | Min | 0.281 | 0.066 | 0.181 | -0.028 | 0.141 | -0.038 |
| Few-zero ($\beta_0 = 0.5$) | Max | 0.930 | 0.890 | 0.714 | 0.721 | 0.627 | 0.607 |
| | 75-percentile | 0.842 | 0.742 | 0.548 | 0.519 | 0.549 | 0.343 |
| | Median | 0.775 | 0.700 | 0.478 | 0.413 | 0.500 | 0.263 |
| | 25-percentile | 0.683 | 0.596 | 0.389 | 0.273 | 0.457 | 0.162 |
| | Min | 0.270 | 0.088 | 0.156 | 0.027 | 0.374 | -0.074 |

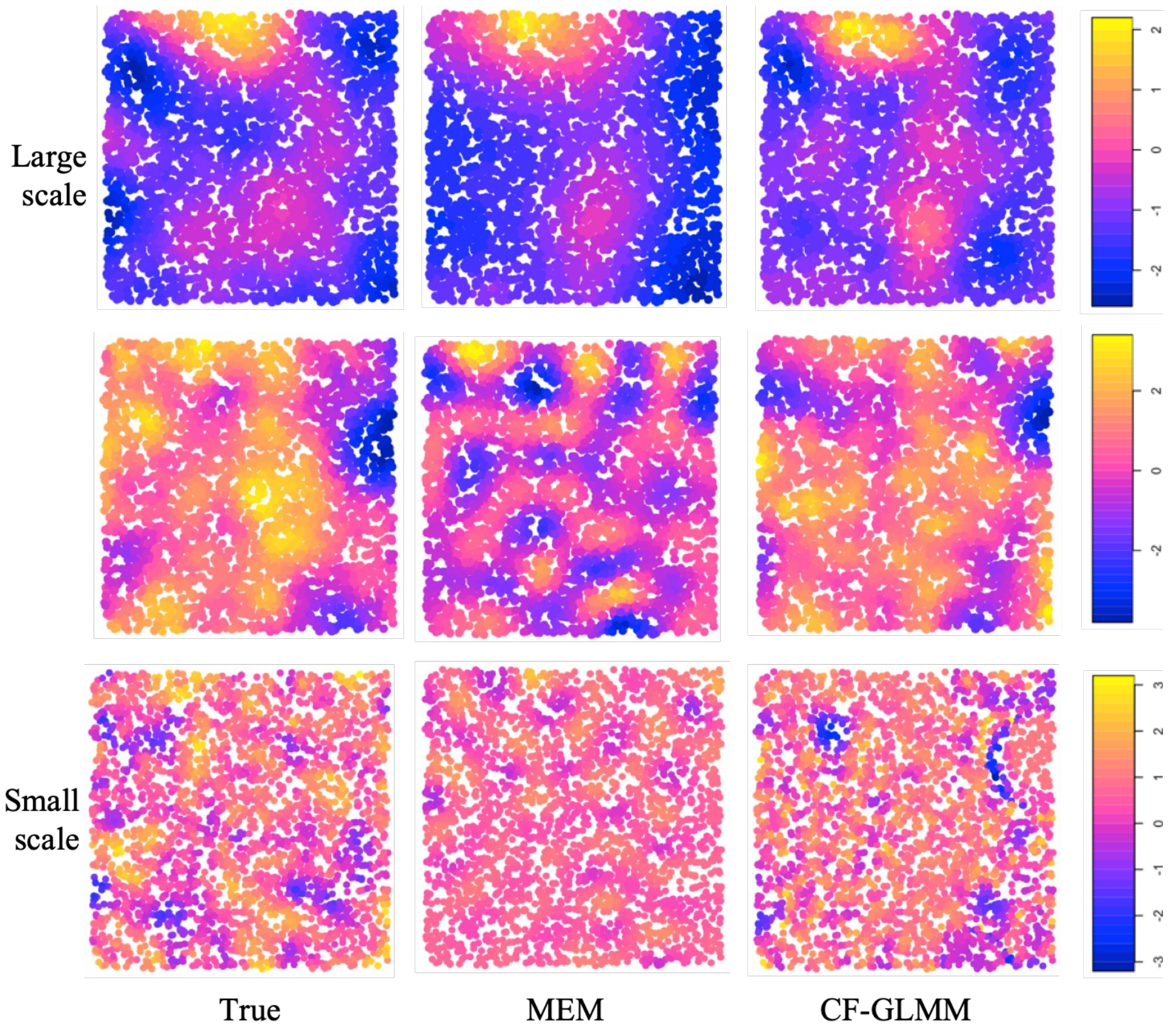

Figure 7: Large-, moderate-, and small-scale components and their estimates

# 6. Application

## 6.1. Outline

This section applies the proposed method to analyze coronavirus disease 2019 (COVID-19) cases in Tokyo Metropolis, Japan. The study periods include an early period (January–May 2020) and a late period (July–December 2021). The early period corresponds to an initial outbreak, characterized by limited testing and strict interventions, whereas the late period corresponds to a major outbreak driven by the Delta variant, marked by high case counts.

The response variable is the number of confirmed infections in each 500-m grid during each period (see Figure 8). The data were constructed by aggregating the facility-level infection counts collected by JX PRESS Corporation. Because the data do not represent confirmed case counts by place of residence, but instead reflect infection events at specific locations, they are suitable for analyzing spatiotemporal patterns of outbreaks, as demonstrated by Nakaya and Nagata (2024). However, the dataset may be noisy because it is primarily based on voluntarily disclosed online information from facilities and organizations. This study aims to elucidate the latent spatial distribution of COVID-19 risk and to identify underlying factors contributing to its spatial variability. Note that, to eliminate grids with no human activity, primarily mountainous and river areas, cells with no reported infection cases over 2020–2021 were excluded from the analysis.

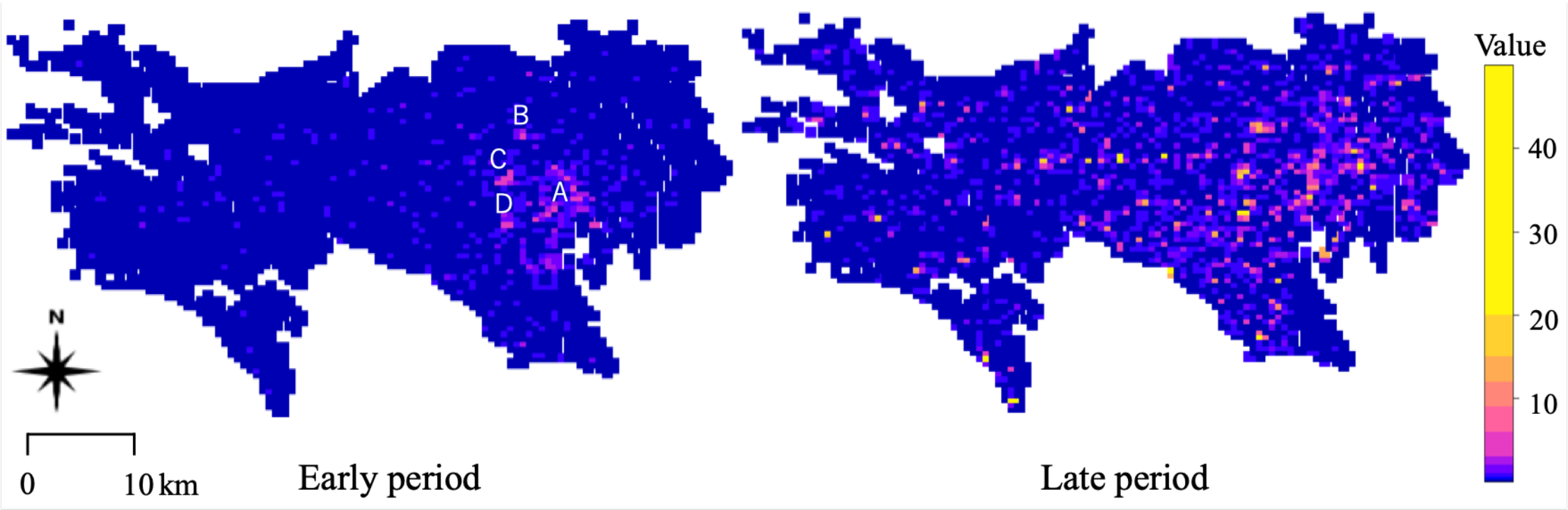

Figure 8: The number of infected cases by 500 m grid in the early and late periods. A–D in the left panel denote (A) the center of Tokyo, and the subcenters including (B) Ikebukuro, (C) Shinjuku, and (D) Shibuya.

The explanatory variables include the average ambient population between 12:00 and 12:59, two weeks prior (Pop; source: NTT DoCoMo Inc.), the distance to the nearest railway station (Station), a dummy variable indicating commercial use areas (Com), and another dummy variable indicating industrial use areas (Ind). The usefulness of Pop for modeling COVID-19 spread has been confirmed in previous studies (e.g., Nagata et al. 2021). The two-week lag in Pop was introduced to reflect the approximate delay between infection and diagnosis or reporting (Nagata et al. 2024).

Data other than Pop were obtained from the National Land Numerical Information download site (https://nlftp.mlit.go.jp/ksj/). The proposed Poisson CF-GLMM is compared with a basic Poisson GLM.

## 6.2.Result

GLM and CF-GLMM were estimated for each period. As summarized in Table 5, both the Poisson deviance and the pseudo-$R^2$ indicate that CF-GLMM achieves better modeling accuracy than GLM, highlighting the importance of accounting for spatial dependence. In the first period, Station is significant in GLM but not in CF-GLMM, whereas the reverse pattern is observed for Ind. Given the superior coefficient estimation accuracy of CF-GLMM demonstrated in Section 4, its results are likely to be more reliable. In both periods, the CF-GLMM coefficients on Pop, Com, and Ind are positive, whereas the coefficients on Station are negative. Based on the results, infection cases tend to increase in commercial areas near railway stations with large ambient populations two weeks prior. This result is consistent with Imamura et al. (2023), who showed that cluster outbreaks in nightlife settings including restaurants, bars, and nightclubs, most of which are located in commercial areas near railway stations, have a strong influence on the spread of infections. Infection cases are also higher in industrial areas, likely owing to contact opportunities within workplaces.

Table 5: Estimated coefficients, deviance, and McFadden's pseudo-$R^2$ (McFadden 1972)

| | Early period | | | | Late period | | | |
|---|---|---|---|---|---|---|---|---|
| | GLM | | CF-GLMM | | GLM | | CF-GLMM | |
| | Est. | SE[1] | Est. | SE | Est. | SE | Est. | SE |
| Intercept | -21.12 | 0.944 ***[2] | -17.10 | 0.924 *** | -12.12 | 0.323 *** | -18.34 | 0.315 *** |
| Pop | 1.893 | 0.088 *** | 1.451 | 0.086 *** | 1.175 | 0.030 *** | 1.820 | 0.030 *** |
| Station | -0.335 | 0.163 ** | -0.251 | 0.164 | -0.182 | 0.055 *** | -0.334 | 0.052 *** |
| Com | 0.332 | 0.132 ** | 0.450 | 0.132 *** | 0.252 | 0.057 *** | 0.284 | 0.057 *** |
| Ind | 0.049 | 0.121 | 0.287 | 0.117 ** | 0.483 | 0.050 *** | 0.405 | 0.050 *** |
| Deviance | 1580 | | 1400 | | 6678 | | 5497 | |
| Pseudo-$R^2$ | 0.363 | | 0.522 | | 0.279 | | 0.580 | |

[1] SE stands for standard error

[2] *** and ** indicate statistical significance at the 1 % and 5 % levels, respectively, based on z-tests.

Figure 9 shows the predicted infection counts, which can be interpreted as an index of infection risk. Both the GLM and CF-GLMM indicate persistent hotspots in central Tokyo and major subcenters, including Shinjuku, Ikebukuro, and Shibuya (see Figure 8). Compared with the GLM, the CF-GLMM predictions exhibit more pronounced stripe patterns along railway lines, many of which radiate from central Tokyo. As a result, under the CF-GLMM, local hotspots are distributed along railway lines, particularly in the late period, whereas they are relatively limited in the early period. This contrast is less evident in the GLM results, suggesting that CF-GLMM is better suited to capturing localized spatial patterns.

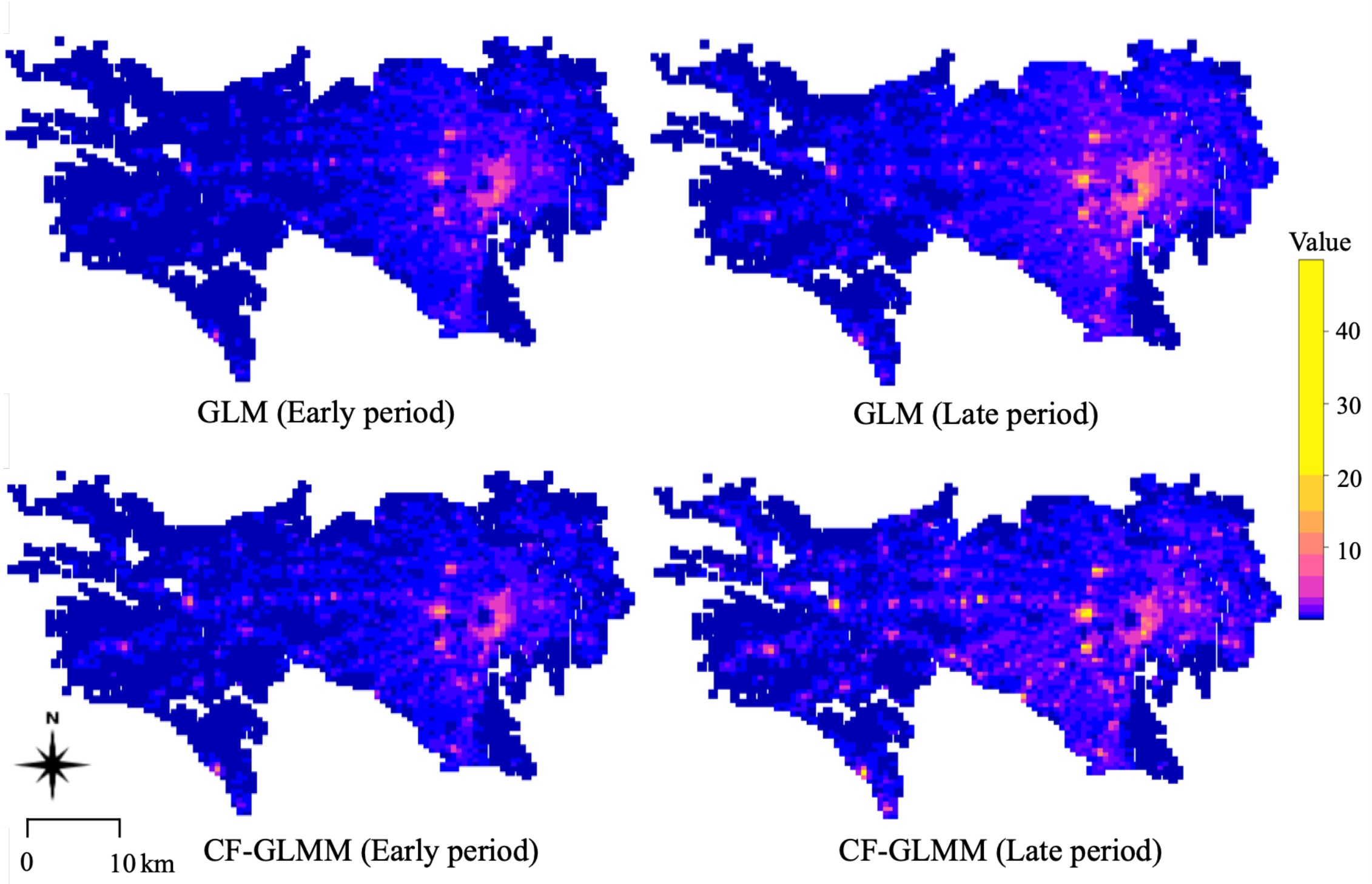

Figure 9: Predicted infected cases

Figure 10 plots the coefficient of variation (CoV), defined as the ratio of the predictive standard deviation to the predictive mean. A larger CoV indicates greater relative uncertainty in the predictions. As shown in the figure, GLM exhibits smaller CoV values, likely due to underestimation of uncertainty arising from the neglect of spatial dependence (see Lee 2011). In contrast, CF-GLMM addresses this issue and yields larger uncertainty estimates that better account for spatial dependence. The CoV values tend to increase in the western suburban areas, where infection cases are relatively scarce, and this tendency is more pronounced in the early period, when case counts are even lower.

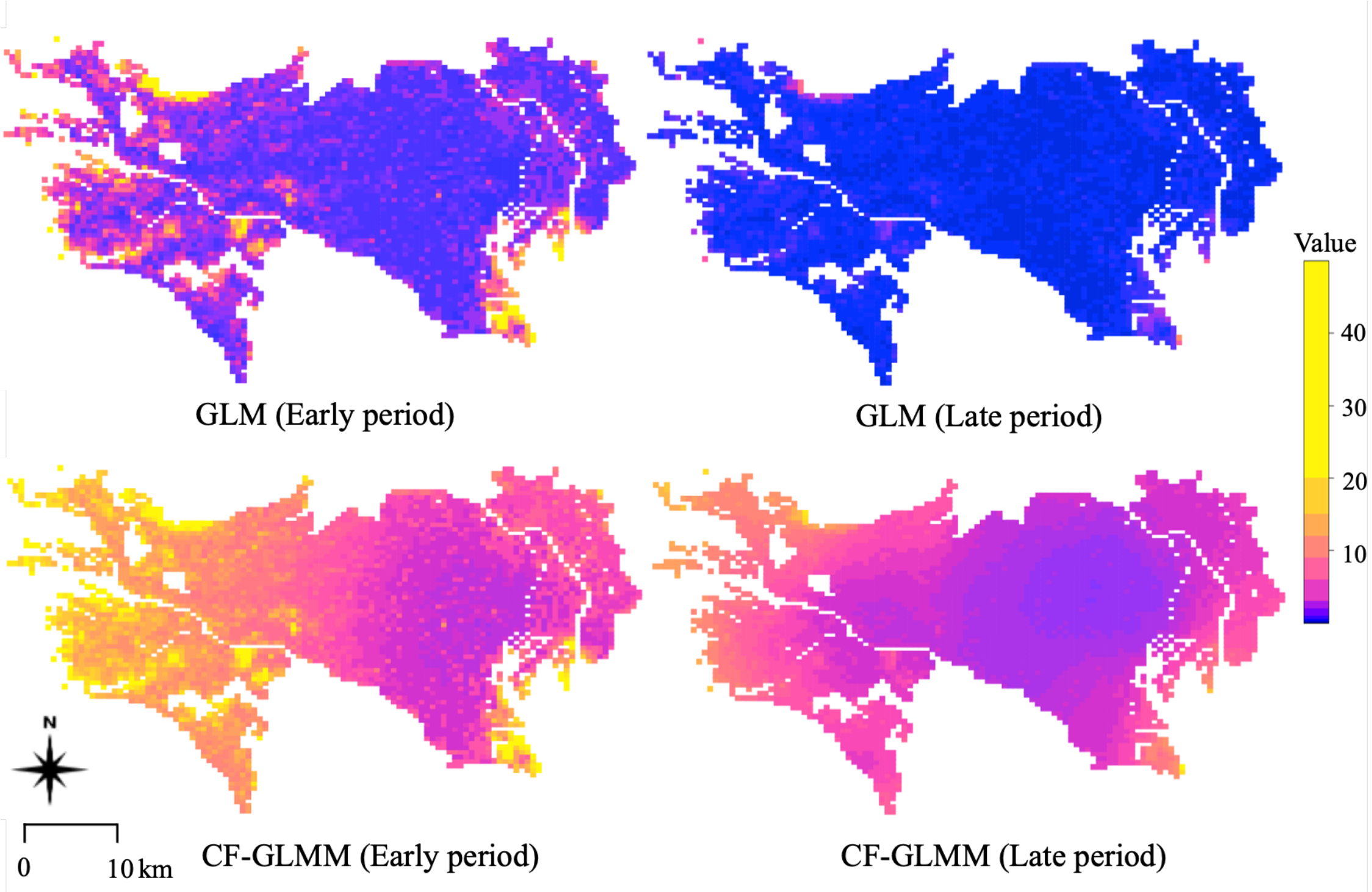

Figure 10: Predicted coefficient of variation

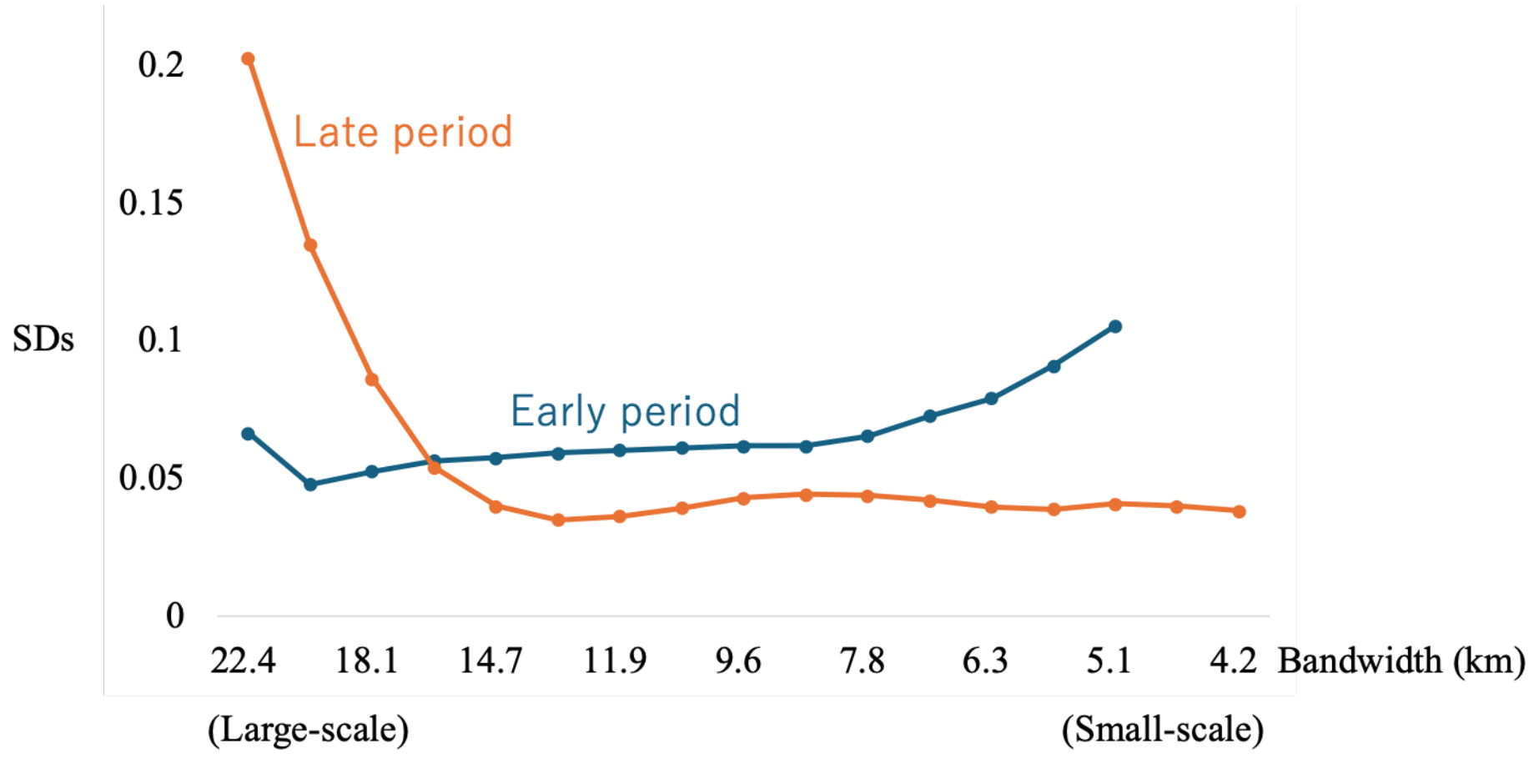

Figure 11: Standard deviations of each scale-wise process.

Figure 11 plots the standard deviations (SDs) of each scale-wise spatial process that contributes to the predictive values. Larger SDs indicate a greater contribution of the corresponding scale-specific process to spatial variation. Interestingly, smaller-scale processes corresponding to smaller bandwidths have larger contributions in the early period, whereas the opposite is observed in the late period. These results clearly suggest a shift from an early spatial pattern concentrated around central Tokyo to a more dispersed pattern.

Finally, the scale-wise processes are divided into a large-scale component corresponding to $h_r \geq 10$ km and a small-scale component corresponding to $h_r < 10$ km, and plotted in Figure 12. These components adjust the baseline infection risk predicted by the covariates based on large- and small-scale residual patterns.

In the early period, both components exhibit elevated risks near the city center, indicating that infections were largely concentrated in central areas. In the late period, the large-scale component suggests increased risk in the western suburban areas, while the small-scale component indicates elevated risk in the western and eastern parts of the center, which are characterized by dense residential areas. These results are consistent with Imamura et al. (2023), who reported an increase in household transmission during the late period in Tokyo.

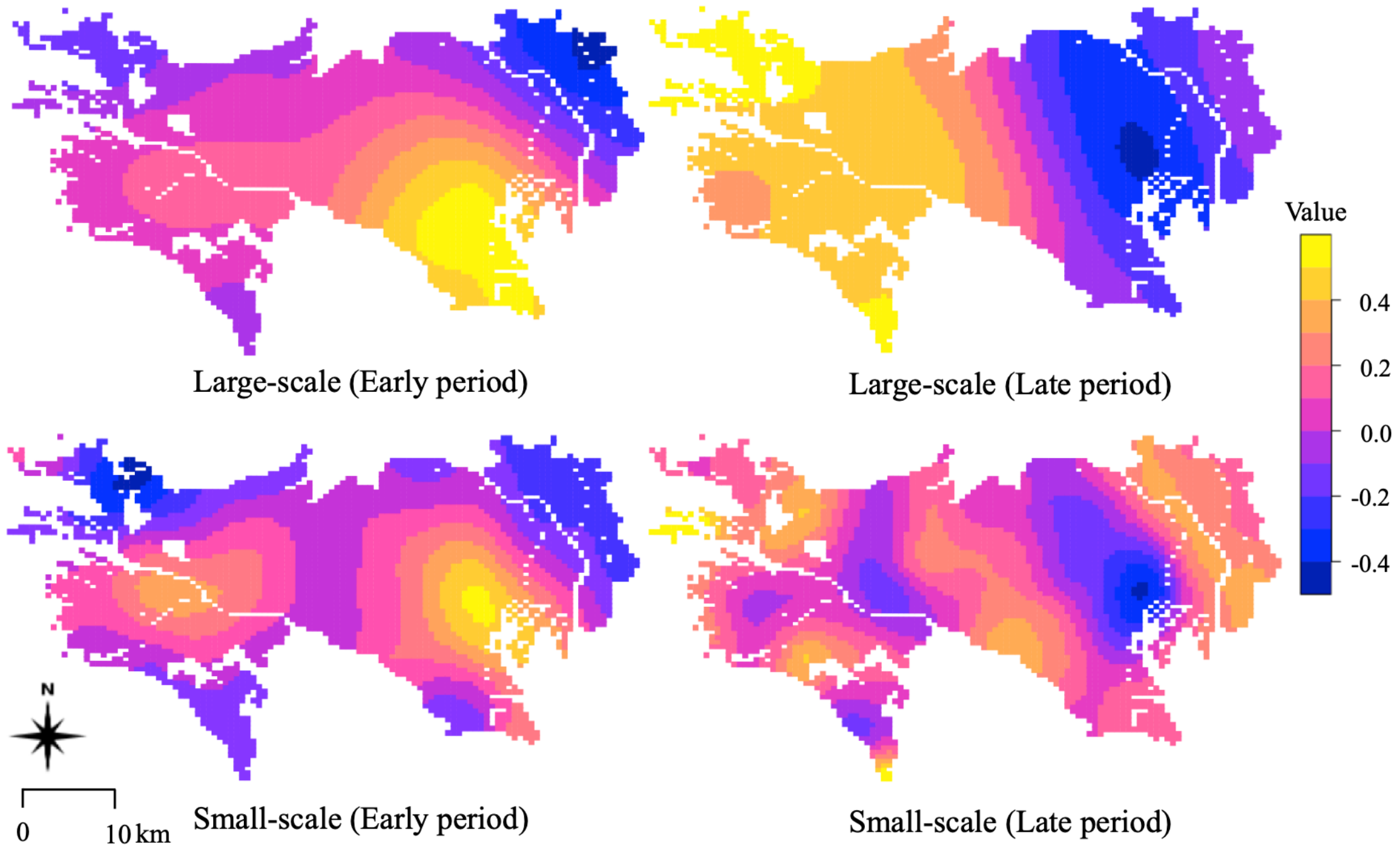

Figure 12: Large- and small-scale processes

## 7. Concluding remarks

This study extends CFSM, originally developed for Gaussian data, to exponential-family responses through CF-GLMM. Whereas many spatial GLMMs represent latent processes through covariance modeling, the proposed method constructs the latent process by aggregating local models, offering an alternative perspective.

Although the proposed method can be expressed as a basis-function model, it adopts a distinctive validation-guided coarse-to-fine construction. The terminal number $L$ of basis functions need not be specified a priori but is determined through the HV-based selection of $R$. This construction reduces the risk of oversmoothing caused by an insufficiently rich basis representation. The simulation experiments and empirical application demonstrated favorable predictive accuracy and computational efficiency.

Nevertheless, several limitations warrant further investigation. The theoretical properties of CF-GLMM, including consistency, convergence, identifiability of the multiscale decomposition, and uncertainty associated with the HV-based selection of $R$, remain to be established. In addition, the ability of CF-GLMM to handle more complex spatial structures, such as anisotropy and non-stationarity, has not yet been systematically evaluated. Finally, the empirical comparison was restricted to spatial models with readily available implementations for non-Gaussian responses. Further work should compare CF-GLMM more broadly with other scalable approaches, including Vecchia-type (Katzfuss and Guinness 2021), nearest-neighbor (Datta et al. 2016), and multiresolution approximations (Nychka et al. 2015).

Potential future directions include spatially varying coefficient modeling (e.g., Wolf et al. 2018; Comber et al. 2023; Murakami et al. 2024), multivariate modeling (e.g., Gelfand 2021), and distributional regression modeling (Klein et al. 2015). By replacing the static local model with a dynamic one, the proposed method can be extended to incorporate local temporal dynamics, enabling flexible spatiotemporal modeling. In addition, extending the model to allow the weighting kernels and number of scales $R$ to vary spatially is likely to enhance the capability of CF-GLMM for modeling complex spatial processes, including the anisotropic and non-stationary processes noted above.

The proposed method can be viewed as an extractor of multiscale spatial basis functions. These basis functions can be incorporated into modern machine-learning models, such as random forests and deep neural networks, to potentially improve spatial prediction. Various spatial features, including kernels, kriging predictors, and raw geographic coordinates, have been explored in this context (see Kopczewska 2022). The proposed multiscale, data-adaptive basis functions provide another candidate representation whose incremental benefit in downstream machine-learning models should be evaluated in future work.

There remains room for further improvement in computational efficiency. For example, in the HV-based optimization of $R$, subsamples may be sufficient for tuning. In addition, the estimation of the local models is naturally parallelizable. Moreover, because the method relies primarily on relatively simple operations, it may also be amenable to GPU-based implementations. Such sampling and parallel-computing schemes are expected to be particularly effective for very large datasets containing hundreds of thousands or millions of observations.

Overall, the proposed framework provides a scalable and flexible alternative to covariance-based spatial GLMMs, combining accurate prediction, data-adaptive spatial refinement, exploratory multiscale analysis, and broad extensibility.

## Data availability statement

CF-GLMM is implemented in the R package spCF (https://cran.r-project.org/web/packages/spCF/). The data used in the COVID-19 study were obtained from a commercial data provider and are subject to licensing restrictions. They are therefore not publicly available.

# Appendix 1. Details of the uncertainty quantification

This appendix details the empirical variance calibration and the cluster-robust coefficient covariance summarized in Section 3.4. Throughout, the holdout validation (Section 3.3) splits the samples into $N_t$ training and $N_v$ validation sites, and the weighted squared loss (Eq. 9) is minimized by fitting the following working model:

$$\hat{\eta}(s_{i_t}) = \mathbf{x}(s_{i_t})'\widehat{\boldsymbol{\beta}} + o(s_{i_t}) + \hat{z}_{1:R}(s_{i_t}) + e(s_{i_t}), \qquad e(s_{i_t}) \sim N\left(0, v^2 \widehat{w}_y^{-1}(s_{i_t})\right). \tag{A1}$$

The residual is given by $\hat{e}(s_{i_t}) = \hat{\eta}(s_{i_t}) - \mathbf{x}(s_{i_t})'\widehat{\boldsymbol{\beta}} - o(s_{i_t}) - \hat{z}_{1:R}(s_{i_t})$. Since the spatially dependent process $\hat{z}_{1:R}(s_i)$ has been fitted at the training sites, the residual variance $\hat{v}^2 = \frac{\sum_{i_t} \widehat{w}_y(s_{i_t})\, \hat{e}(s_{i_t})^2}{\sum_{i_t} \widehat{w}_y(s_{i_t})}$ approximates the variance of the independent noise.

## A1.1. Empirical calibration of the process variance

The model-based process variance $\hat{\sigma}^2_{1:R}(s_i) = \sum_{r=1}^{R} \hat{\sigma}^2_r(s_i)$, which may be biased, is rescaled so that (A) its mean matches (B) the empirical process variance $\hat{\sigma}^2_{emp}$. Given the working weight $\widehat{w}_y(s_{i_v})$, (A) the mean of $\hat{\sigma}^2_{1:R}(s_i)$ yields $\bar{\sigma}^2_{1:R} = \frac{\sum_{i_v} \widehat{w}_y(s_{i_v})\, \hat{\sigma}^2_{1:R}(s_{i_v})}{\sum_{i_v} \widehat{w}_y(s_{i_v})}$. To evaluate (B), we first compute the residual variance for the validation samples. Analogous to $\hat{v}^2$, it is obtained as $\hat{v}^2_{emp} = \frac{\sum_{i_v} \widehat{w}_y(s_{i_v})\, \hat{e}(s_{i_v})^2}{\sum_{i_v} \widehat{w}_y(s_{i_v})}$ where $\hat{e}(s_{i_v}) = \hat{\eta}(s_{i_v}) - \mathbf{x}(s_{i_v})'\widehat{\boldsymbol{\beta}} - o(s_{i_v}) - \hat{z}_{1:R}(s_{i_v})$. Since $\hat{z}_{1:R}(s_{i_v})$ is predicted at validation sites rather than fitted, the residual variance $\hat{v}^2_{emp}$ reflects the prediction error $z_{1:R}(s_i) - \hat{z}_{1:R}(s_i)$ as well as the independent noise. Therefore, the process variance can be estimated as $\hat{\sigma}^2_{emp} = \max(\hat{v}^2_{emp} - \hat{v}^2, 0)$, which subtracts independent noise $\hat{v}^2$ from $\hat{v}^2_{emp}$ subject to a non-negativity constraint. By rescaling $\hat{\sigma}^2_{1:R}(s_i)$ so that its mean $\bar{\sigma}^2_{1:R}$ matches $\hat{\sigma}^2_{emp}$, we obtain $\hat{\sigma}^{*2}_{1:R}(s_i) = \frac{\hat{\sigma}^2_{emp}}{\bar{\sigma}^2_{1:R}} \hat{\sigma}^2_{1:R}(s_i)$.

## A1.2. Cluster-robust coefficient covariance

Because the fitted process $\hat{z}_{1:R}(s_i)$ is fixed when estimating $\boldsymbol{\beta}$, the model-based covariance understates $Var(\hat{\boldsymbol{\beta}})$. Therefore, we use a spatial-block cluster-robust (sandwich) estimator (Bester et al. 2011),

$$\hat{\boldsymbol{\Sigma}}_{\boldsymbol{\beta}} = (\mathbf{X}'\mathbf{W}\mathbf{X})^{-1}(\mathbf{B}_e + \mathbf{B}_z)(\mathbf{X}'\mathbf{W}\mathbf{X})^{-1} \quad \text{(A2)}$$

where $\mathbf{X} = [\mathbf{x}(s_1), \ldots, \mathbf{x}(s_N)]'$, $\mathbf{W} = \text{diag}(w_y(s_1), \ldots, w_y(s_N))$. The meat splits into an observation-noise component and a spatial process component. The noise component is $\mathbf{B}_e = \frac{G}{G-1} \sum_g \mathbf{q}_g \mathbf{q}_g'$ with block score sums $\mathbf{q}_g = \sum_{i \in g} w_y(s_i) \left(\hat{\eta}(s_i) - \mathbf{x}(s_i)'\hat{\boldsymbol{\beta}} - o(s_i)\right) \mathbf{x}(s_i)$, where $G$ is the number of blocks. The process component adds the process uncertainty back through its calibrated variance, $\mathbf{B}_z = \sum_g \sum_{i,j \in g} \rho(d_{ij}) \mathbf{u}(s_i)\mathbf{u}(s_j)'$, where $\mathbf{u}(s_i) = w_y(s_i)\, \sigma^*_{1:R}(s_i)\, \mathbf{x}(s_i)$, $\hat{\sigma}^*_{1:R}(s_i) = \sqrt{\hat{\sigma}^{*2}_{1:R}(s_i)}$ is the calibrated process standard deviation from Appendix A1.1. Here, $d_{ij}$ is the distance between sites $i$ and $j$, and $\rho(d_{ij}) = \exp(-d_{ij}/h_{med})$ is a within-block correlation whose range $h_{med}$ is the median of the bandwidths selected in the HV. Modeling the field as a smooth correlated error through its calibrated variance, rather than retaining its realized value in the residual, yields near-nominal coverage while avoiding the conservatism of the field-retained estimator. The blocks form a $G_x \times G_y$ grid obtained by cutting each coordinate axis at its empirical quantiles, with per-axis counts chosen as $G_x = \begin{cases} 2 & \text{if } G_x^0 < 2 \\ G_x^0 & \text{if } 2 \leq G_x^0 \leq 8 \\ 8 & \text{if } 8 < G_x^0 \end{cases}$ where $G_x^0 = \text{round}(L_x/\, h^{med})$. $L_x$ is the horizontal extent of the coordinates and $G_y$ is defined analogously from the vertical extent $L_y$. The bounds $[2,8]$ keep the number of blocks neither too small for stable estimation nor so large that the blocks become narrower than the dependence range, which would leave neighboring blocks correlated.

For count responses the working weight $w_y(s_i)$ grows with the fitted mean, so $\mathbf{B}_z$ can overestimate the coefficient variance at sites where the process is large but well estimated. We therefore cap $\hat{\boldsymbol{\Sigma}}_{\boldsymbol{\beta}}$ at a refit-free leave-one-out (LOO) estimate. Because the fitted process is locally a linear smoother of the working response, the LOO block score inflates each residual, $\mathbf{q}_g^{loo} = \sum_{i \in g} w_y\,(s_i)\,\frac{\hat{\eta}(s_i) - \mathbf{x}(s_i)'\hat{\boldsymbol{\beta}} - o(s_i)}{1 - h_{ii}}\,\mathbf{x}(s_i)$, and $\mathbf{B}_{loo} = \frac{G}{G-1}\sum_g \mathbf{q}_g^{loo}\,\mathbf{q}_g^{loo\prime}$. The leverage $h_{ii} = h_i^x + h_i^z$ combines the fixed-effect leverage $h_i^x = w_y(s_i)\,\mathbf{x}(s_i)'(\mathbf{X}'\mathbf{W}\mathbf{X})^{-1}\mathbf{x}(s_i)$ and the process leverage $h_i^z = 1 - \prod_r \left[1 - \frac{1}{\sum_j \exp(-d_{ij}/h_r)}\right]$ aggregated over the selected bandwidths $h_r$. Writing $\hat{\boldsymbol{\Sigma}}_{\boldsymbol{\beta}}^e =$ $(\mathbf{X}'\mathbf{W}\mathbf{X})^{-1}\mathbf{B}_e(\mathbf{X}'\mathbf{W}\mathbf{X})^{-1}$ and $\hat{\boldsymbol{\Sigma}}_{\boldsymbol{\beta}}^{loo} = (\mathbf{X}'\mathbf{W}\mathbf{X})^{-1}\mathbf{B}_{loo}(\mathbf{X}'\mathbf{W}\mathbf{X})^{-1}$, we cap each coefficient variance at the LOO estimate, floored at the noise-only value, $\text{Var}(\hat{\beta}_k) = \max\{[\hat{\boldsymbol{\Sigma}}_{\boldsymbol{\beta}}^{e}]_{kk},\ \min([\hat{\boldsymbol{\Sigma}}_{\boldsymbol{\beta}}]_{kk}, [\hat{\boldsymbol{\Sigma}}_{\boldsymbol{\beta}}^{loo}]_{kk})\}$, and rescale $\hat{\boldsymbol{\Sigma}}_{\boldsymbol{\beta}}$ to this diagonal while preserving its correlation. The cap is self-calibrating: it binds only where the process term overstates the variance (count responses) and typically has no effect on Gaussian and binomial responses, for which the estimator is already calibrated.

# Appendix 2. Spatial covariance under an idealized convolution model

This appendix derives the covariance of an idealized process-convolution model motivated by the multiscale construction of CF-GLMM. We assume independent, mean-zero Gaussian white-noise measures $W_r$ across scales, with variance $\tau_r^2$, and convolution over $\mathbb{R}^D$, where $D = 2$. To relate this model to discrete local smoothing, we assume approximately uniform sampling and disregard boundary effects.

Let $z_r(s) = \int_{\mathbb{R}^D} k((s-u)/h_r)\, dW_r(u)$ with the exponential kernel $k(u) = \exp(-||u||)$ used in CF-GLMM, where $||u||$ denotes the Euclidean norm. This is an idealized continuous-space analogue of the single-scale process in Eq. (2), in which the gPoE-based kernel averaging of local means is replaced by kernel averaging of white noise. The covariance is $\text{Cov}(z_r(s), z_r(s')) = \tau_r^2 \int_{\mathbb{R}^D} k((s-u)/h_r) k((s'-u)/h_r) du$. A change of variables and translation invariance therefore yield $\mathrm{Cov}(z_r(s), z_r(s')) = \tau_r^2 h_r^D (k * k)((s - s')/h_r) = \sigma_r^2 \rho_0(d/h_r)$, where $d = \|s - s'\|$, $\sigma_r^2 = \tau_r^2 h_r^D (k * k)(0)$, and $\rho_0(x) = (k * k)(x)/(k * k)(0)$. Here $*$ denotes convolution.

To obtain $\rho_0(\cdot)$ explicitly, we use the convolution theorem, which gives $\mathcal{F}[k * k] = |F[k]|^2$ for the symmetric kernel $k$. Under the Fourier convention $F[k](\omega) = \int_{\mathbb{R}^D} \exp(-i\omega \cdot u) k(u) du$, the exponential kernel has $\mathcal{F}[k](\omega) \propto (1 + \|\omega\|^2)^{-(D+1)/2}$. Hence, the spectral density of $\rho_0(\cdot)$ has shape $|\mathcal{F}[k](\omega)|^2 \propto (1 + \|\omega\|^2)^{-(D+1)}$. Comparing this with the Matérn spectral shape $(\kappa^2 + \|\omega\|^2)^{-(\nu + D/2)}$ identifies $\nu = D/2 + 1$. In two dimensions, each idealized single-scale process $z_r(s)$ has a Matérn covariance with $\nu = 2$ and inverse range $\kappa_r = 1/h_r$; its normalized form is $\rho_0(x) = x^2 K_2(x)/2$ for $x > 0$ and $\rho_0(0) = 1$, where $K_2$ is the modified Bessel function of the second kind.

Independence among the scale-wise processes gives the exact covariance $C(d) = \sum_r \sigma_r^2 \rho_0(d/h_r)$ for their finite sum. This is a mixture of Matérn covariances with a common smoothness and different ranges, and is not generally a single Matérn covariance. This result provides a theoretical reference for interpreting the multiscale structure of CF-GLMM, rather than an exact characterization of the fitted model.

# Appendix 3. Computational complexity

We analyze the cost of one coarse-to-fine sweep; the holdout tuning that selects the number of scales and the full-sample fit share the same structure. Assume the *N* observations have an approximately uniform density over a fixed spatial domain of area *A*. At scale *r* with bandwidth $h_r$, the process is fitted by local models anchored at knots, whose number is $n_r \propto A/h_r^2$ (capped at the number of distinct locations), each using the observations within a radius proportional to $h_r$, i.e. $m_r \propto (N/A)\, h_r^2$ neighbors. The per-scale cost is therefore $n_r m_r \propto \frac{A}{h_r^2} \cdot \frac{N h_r^2}{A} = O(N)$, independent of $h_r$: coarse scales fit few models over many neighbors, whereas fine scales fit many models over few neighbors, so their product stays proportional to *N*. Note that the k-d tree algorithm for the k-means clustering adds only a logarithmic factor to the neighbor queries, and at fine scales the knots are chosen by subsampling to avoid a quadratic k-means cost.

The bandwidths decay following $h_r = \delta h_{r-1}$ with $0 < \delta < 1$, from a coarse initial value down to a floor $h_{\min}$ set at about half the median nearest-neighbor distance; under uniform density this distance is $O(N^{-1/2})$, so the number of scales is $R = O(\log(h_1/h_{\min})) = O(\log N)$.

Combining the two, a single sweep costs O(*N*) × O(log *N*) = O(*N* log *N*), and its peak memory is *O*(*N*), because the neighbor lists are streamed in fixed-size chunks. This contrasts with the $O(N^3)$ time and $O(N^2)$ memory of dense covariance-based GP/GLMMs. The near-linear behavior is consistent with the computation-time results (Table 2), which grow only marginally faster than linearly with *N*; the local-model fits are, moreover, embarrassingly parallel over knots.